\documentclass[10pt,twocolumn,journal]{IEEEtran}
\usepackage{times}
\usepackage{amsbsy}
\usepackage{amsmath}
\usepackage{latexsym}
\usepackage{amssymb}
\usepackage[ansinew]{inputenc}
\usepackage[dvips ]{graphicx}
\usepackage[dvips]{pstcol}
\input epsf

\title{Blind Adaptive Constrained Constant-Modulus Reduced-Rank
Interference Suppression Algorithms Based on Interpolation, Switched
Decimation and Filtering \vspace{-0.05em}}
\author{Rodrigo C. de Lamare, Raimundo Sampaio-Neto and Martin
Haardt\vspace{-0.05em}
\thanks{Copyright (c) 2010 IEEE.
Personal use of this material is permitted. However, permission to
use this material for any other purposes must be obtained from the
IEEE by sending a request to pubs-permissions@ieee.org. This work
was partially funded by the Ministry of Defence (MoD), UK,
Contract No. RT/COM/S/021. Dr. R. C. de Lamare is with the
Department of Electronics, University of York, York Y010 5DD,
United Kingdom, Prof. R. Sampaio-Neto is with CETUC/PUC-RIO,
22453-900, Rio de Janeiro, Brazil and Prof. Haardt is with the
Communications Research Laboratory, Ilmenau University of
Technology, Germany. E-mails: rcdl500@ohm.york.ac.uk,
raimundo@cetuc.puc-rio.br and Martin.Haardt@tu-ilmenau.de}}

\begin{document}

\maketitle
\begin{abstract}{
This work proposes a blind adaptive reduced-rank scheme and
constrained constant-modulus (CCM) adaptive algorithms for
interference suppression in wireless communications systems. The
proposed scheme and algorithms are based on a two-stage processing
framework that consists of a transformation matrix that performs
dimensionality reduction followed by a reduced-rank estimator. The
complex structure of the transformation matrix of existing methods
motivates the development of a blind adaptive reduced-rank
constrained (BARC) scheme along with a low-complexity reduced-rank
decomposition. The proposed BARC scheme and a reduced-rank
decomposition based on the concept of joint interpolation,
switched decimation and reduced-rank estimation subject to a set
of constraints are then detailed. The proposed set of constraints
ensures that the multi-path components of the channel are combined
prior to dimensionality reduction. In order to cost-effectively
design the BARC scheme, we develop low-complexity decimation
techniques, stochastic gradient and recursive least squares
reduced-rank estimation algorithms. A model-order selection
algorithm for adjusting the length of the estimators is devised
along with techniques for determining the required number of
switching branches to attain a predefined performance. An analysis
of the convergence properties and issues of the proposed
optimization and algorithms is carried out, and the key features
of the optimization problem are discussed. We consider the
application of the proposed algorithms to interference suppression
in DS-CDMA systems. The results show that the proposed algorithms
outperform the best known reduced-rank schemes, while requiring
lower complexity.

}
\end{abstract}

\section{Introduction}

Interference suppression in wireless communications has attracted
a great deal of attention in the last decades
\cite{honig&poor,verdu}. Motivated by the need to counteract the
effects of wireless channels, to increase the capacity of multiple
access schemes, and to enhance the quality of wireless links, a
plethora of schemes and algorithms have been proposed for
equalization, multiuser detection and beamforming. These
techniques have been applied to a variety of standards that
include spread spectrum \cite{honig&tsatsanis}, orthogonal
frequency-division multiplexing (OFDM) \cite{stuber} and
multi-input multi-output (MIMO) systems \cite{vblast} and continue
to play a key role in the design of wireless communications
systems.

\subsection{Prior Work}

{  In order to design interference mitigation techniques,
designers are required to employ estimation algorithms for
computing the parameters of the filters used at the receiver or at
the transmitter. In the literature of estimation algorithms, one
can broadly divide them into supervised and blind techniques.
Blind methods are appealing because they can alleviate the need
for training sequences or pilots, thereby increasing the
throughput and efficiency of wireless networks. In particular,
blind estimation algorithms based on constrained optimization
techniques are important in several areas of signal processing and
communications such as beamforming and interference suppression
\cite{haykin}. The constrained optimizations in these applications
usually deal with linear constraints that correspond to prior
knowledge of certain parameters such as direction of arrival (DoA)
of users' signals in antenna-array processing \cite{liberti} and
the signature sequence of the desired signal in CDMA interference
suppression \cite{honig}. Numerous blind estimation algorithms
with different trade-offs between performance and complexity have
been reported in the last decades
\cite{honig}-\cite{delamareccmmimo}.} {The designs based on the
constrained constant modulus (CCM) criterion
\cite{kwak,xu&liu,delamareccm,mswfccm,delamareccmmimo} have shown
increased robustness against signature mismatch and improved
performance over constrained minimum variance (CMV) approaches
\cite{honig,xutsa,delamaretsp}. In general, the convergence and
tracking performances of these algorithms depend on the eigenvalue
spread of the $M\times M$ full-rank covariance matrix
${\boldsymbol R}$ of the input data vector ${\boldsymbol r}[i]$
that contains $M$ samples of the signal to be processed, and the
number of elements $M$ in the estimator \cite{haykin}. When $M$ is
large, blind algorithms require a large number of samples to reach
their steady-state behavior and may encounter problems in tracking
the desired signal.

Reduced-rank signal processing is a key technique in low-sample
support situations and large optimization problems that has gained
considerable attention in the last few years
\cite{scharf}-\cite{jioccm}. The fundamental idea is to devise a
transformation in such a way that the data vector ${\boldsymbol
r}[i]$ can be represented by a reduced number of effective
features and yet retain most of its intrinsic information content
\cite{scharf}. The goal is to find the best tradeoff between model
bias and variance in a cost-effective way. Prior work on
reduced-rank parameter estimation has considered
eigen-decomposition techniques \cite{wang&poor}, the multi-stage
Wiener filter (MSWF) \cite{goldstein,mswfccm} that is a Krylov
subspace method, the auxiliary vector filtering (AVF) algorithm
\cite{avf5,avf6,avf7,avf8}, the joint and iterative optimization
(JIO) strategy \cite{delamarespl07,delamaretvt10,jioccm} and
adaptive interpolated filters
\cite{delamarecl,delamaresp,delamaretvt}. A major problem with the
MSWF, the AVF-based and the JIO schemes is their high complexity.
Prior work on adaptive interpolated filters
\cite{delamarecl,delamaresp,delamaretvt} has considered MMSE- and
CMV-based designs and shown a significant performance degradation
for rank reduction with large compression ratios. This problem has
been recently addressed by the joint interpolation, decimation and
filtering (JIDF) scheme \cite{jidf_icassp,jidf} for supervised
training. With the exception of the CCM-based MSWF of
\cite{mswfccm} and the JIO of \cite{jioccm}, there is no blind
reduced-rank that has low complexity, good performance and
robustness against signature mismatches.

\subsection{Contributions of This Work}

In this work, we present a low-complexity blind adaptive
reduced-rank constrained scheme (BARC) based on the CCM criterion
and a reduced-rank decomposition using joint interpolation,
switched decimation and reduced-rank estimation. The proposed
scheme is simple, flexible, and provides a substantial performance
advantage over prior art. Unlike the JIDF scheme \cite{jidf}, the
BARC uses an iterative procedure in which the interpolation,
decimation and estimation tasks are jointly optimized using the
CCM design criterion. In the BARC system, the number of elements
for estimation is substantially reduced in comparison with
existing full-rank and reduced-rank schemes, resulting in
considerable computational savings and improved convergence and
tracking performances. A unique feature of the BARC and the
proposed algorithms is that, unlike existing blind schemes, they
do not rely on the full-rank covariance matrix ${\boldsymbol R}$
for performing dimensionality reduction. The BARC and proposed
algorithms skip the processing stage with ${\boldsymbol R}$ and
directly obtain the subspace of interest and constraints via a set
of simple interpolation, decimation and reduced-rank estimation
operations, which leads to much faster convergence and improved
performance. We develop low-complexity decimation techniques,
stochastic gradient (SG) and recursive least squares (RLS)
reduced-rank estimation algorithms. Differently from \cite{jidf},
these algorithms are designed with a set of constraints that are
alternated in the optimization procedure. A model-order selection
algorithm for adjusting the length of the filters is devised along
with techniques for determining the required number of switching
branches to attain a predefined performance. The proposed
model-order selection differs from \cite{jidf} as it employs an
extended filter approach, which is significantly simpler than the
scheme in \cite{jidf} that uses multiple schemes in parallel. The
algorithms for adjusting the number of branches are based on the
constant modulus criterion as opposed to the mean-squared error
(MSE) criterion employed in \cite{jidf}. An analysis of the
convergence properties and aspects of the proposed optimization
and algorithms is also presented. We apply the proposed BARC and
algorithms to interference suppression in DS-CDMA systems. }

{  This paper is organized as follows. The system model of a
DS-CDMA system and the problem statement are presented in Section
II. Section III is dedicated to the description of the BARC scheme
and the CCM reduced-rank estimators. Section IV is devoted to the
presentation of the blind adaptive SG and RLS estimation
algorithms, adjustment of model-order selection and the number of
switching branches, and their complexity. Section V provides an
analysis and a discussion of the proposed optimization problem.
Section VI presents and discusses the simulation results and
Section VII draws the conclusions.}

\section{System Model and Problem Statement}

Let us consider the uplink of a symbol synchronous DS-CDMA system
with $K$ users, $N$ chips per symbol and $L_{p}$ is the maximum
number of propagation paths in chips. A synchronous model is
assumed for simplicity since it captures most of the features of
asynchronous models with small to moderate delay spreads. The
modulation is assumed to have constant modulus. Let us assume that
the signal has been demodulated at the base station, the channel
is constant during each symbol and the receiver is perfectly
synchronized with the main channel path. The received signal after
filtering by a chip-pulse matched filter and sampled at chip rate
yields an $M$-dimensional received vector at time $i$
\begin{equation}
\begin{split}
{\boldsymbol r}[i] & = \sum_{k=1}^{K} A_{k}[i]  {b}_{k}[i]
{\boldsymbol C}_{k} {\boldsymbol h}_k[i]  +
 {\boldsymbol{\eta}}_k[i] + {\boldsymbol n}[i], \label{recsignal}
\end{split}
\end{equation}
where $M=N+L_{p}-1$, ${\boldsymbol n}[i] = [n_{1}[i]
~\ldots~n_{M}[i]]^{T}$ is the complex Gaussian noise vector with
zero mean and $E[{\boldsymbol n}[i]{\boldsymbol n}^{H}[i]] =
\sigma^{2}{\boldsymbol I}$ whose components are independent and
identically distributed, where $(.)^{T}$ and $(.)^{H}$ denote
transpose and Hermitian transpose, respectively, and $E[.]$ stands
for expected value. The user symbols are denoted by ${b}_{k}[i]$,
the amplitude of user $k$ is $A_{k}$[i], {  the first term in
(\ref{recsignal}) represents the user signals transmitted over
multipath channels including the inter-chip interference (ICI),
and ${\boldsymbol{\eta}}_k[i]$ is the inter-symbol interference
(ISI) for user $k$ from the adjacent symbols}. The signature of
user $k$ is represented by ${\boldsymbol s}_{k} = [a_{k}(1) \ldots
a_{k}(N)]^{T}$, the $M\times L_{p}$ constraint matrix
${\boldsymbol C}_{k}$ that contains one-chip shifted versions of
the signature sequence for user $k$ and the $L_{p}\times 1$ vector
${\boldsymbol h}_k[i]$ with the multipath components are described
by
\begin{equation}
{\boldsymbol C}_{k} = \left[\begin{array}{c c c }
a_{k}(1) &  & {\boldsymbol 0} \\
\vdots & \ddots & a_{k}(1)  \\
a_{k}(N) &  & \vdots \\
{\boldsymbol 0} & \ddots & a_{k}(N)  \\
 \end{array}\right],
 {\boldsymbol h}_k[i]=\left[\begin{array}{c} {h}_{k,0}[i]
\\ \vdots \\ {h}_{k,L_{p}-1}[i]\\  \end{array}\right].
\end{equation}
The multiple access interference (MAI) comes from the
non-orthogonality between the received signature sequences,
whereas the ISI span $L_{s}$ depends on the length of the channel
response and how it is related to the length of the chip sequence.
For { $L_{p}=1,~ L_{s}=1$} (no ISI), for { $1<L_{p}\leq N,
L_{s}=2$}, for { $N <L_{p}\leq 2N, L_{s}=3$}, and so on. This
means that at time instant $i$ we will have ISI coming not only
from the previous $L_s-1$ time instants but also from the next
$L_s-1$ symbols. The linear model in (\ref{recsignal}) can be used
to represent other wireless communications systems including MIMO
and OFDM systems. For example, the user signatures of a DS-CDMA
system are equivalent to the spatial signatures of MIMO system.

{  A reduced-rank interference suppression scheme processes the
received vector ${\boldsymbol r}[i]$ in two stages. The first
stage performs a dimensionality reduction via a decomposition of
${\boldsymbol r}[i]$ into a lower dimensional subspace. The second
stage is carried out by a reduced-rank estimator. The output of a
reduced-rank scheme corresponding to the $i$th time instant is
\begin{equation}
z[i] =  \bar{\boldsymbol w}^H[i] {\boldsymbol S}_D^H[i]
{\boldsymbol r}[i]  = \bar{\boldsymbol w}^{H}[i]\bar{\boldsymbol
r}[i], \label{output}
\end{equation}
where ${\boldsymbol S}_{D}[i]$ is an $M \times D$ decomposition
matrix which performs dimensionality reduction and
$\bar{\boldsymbol w}[i]=[ \bar{w}_1^{[i]}
~\bar{w}_2^{[i]}~\ldots\bar{w}_D^{[i]}]^T$ is the $D \times 1$
parameter vector of the reduced-rank estimator. The basic problem
is how to cost-effectively and blindly design the $M \times D$
matrix ${\boldsymbol S}_D[i]$ that transforms the $M \times 1$
vector ${\boldsymbol r}[i]$ into a $D \times 1$ reduced-rank
vector $\bar{\boldsymbol r}[i]$ using the CM criterion .}

\section{Proposed BARC Scheme}

\begin{figure}[h]
       \vspace*{-1em}\centering
  % figura centralizada
       \hspace*{-2.05em}{\includegraphics[width=10cm, height=4cm]{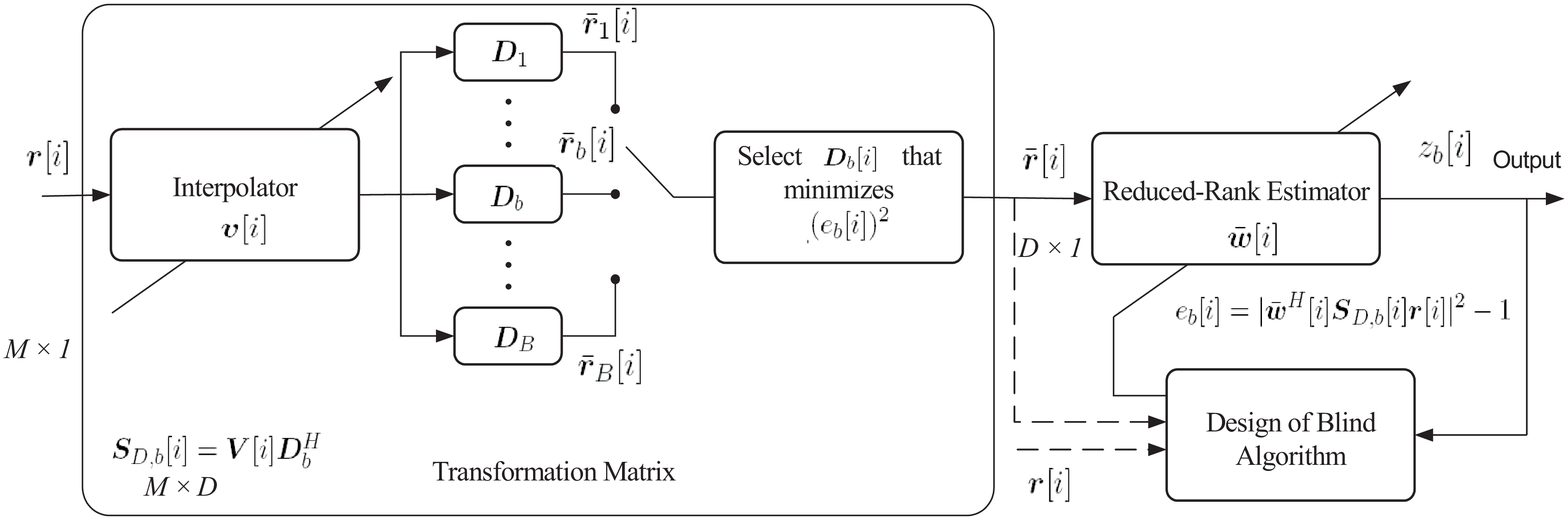}}
        \vspace*{-1em}
       \caption{ Proposed blind adaptive reduced-rank estimation
       structure.}\label{fig1}
       \vspace*{-1.0em}
\end{figure}

{  In this section we introduce the proposed BARC scheme and
detail its key features. The motivation is to improve the
convergence and tracking performance and reduce the complexity.
This is performed via the reduction of the number of coefficients
for computation from $M$ (full-rank schemes) or $D+MD$ (existing
blind reduced-rank schemes) to less than a dozen. The structure of
the BARC scheme is shown in Fig. \ref{fig1}, where an
interpolator, a decimator with several switching decimation
branches and a reduced-rank estimator which are time-varying are
employed.}

{  The $M\times 1$ received vector ${\boldsymbol r}[i]$ is
filtered by the interpolator ${\boldsymbol v}[i] = [v_{0}^{[i]}
\ldots v_{I-1}^{[i]}]^{T}$ with $I$ being the length of the
interpolator and yields the interpolated vector ${\boldsymbol
r}_{\rm I}[i]= {\boldsymbol V}^{H}[i] {\boldsymbol r}[i]$, where
the $M\times M$ convolution matrix ${\boldsymbol V}[i]$ which has
shifted copies of ${\boldsymbol v}[i]$ as described by
\begin{equation}
{\boldsymbol V}[i] =  \left[\hspace*{-0.5em}\begin{array}{c c c c
c c c c c }
v_{0}^{[i]}  & 0 & \ldots  & 0 \\
\vdots & v_{0}^{[i]}  & \ddots  & \vdots \\
v_{ I-1}^{[i]}  & \vdots &  \ldots  & 0 \\
0 & v_{I-1}^{[i]}   & \ddots & 0  \\
\vdots & \vdots  & \vdots  & \vdots \\
0 & 0 &  \ldots & v_{0}^{[i]} \\
 \end{array}\hspace*{-0.5em} \right].
\end{equation}
Let us now express the $M \times 1$ vector ${\boldsymbol r}_{\rm
I}[i]$ in a way that is suitable for algebraic manipulation as a
function of the interpolator ${\boldsymbol v}[i]$:
\begin{equation}
{\boldsymbol r}_{\rm I}[i]={\boldsymbol V}^{H}[i]{\boldsymbol
r}[i] = \boldsymbol{ \Re}_{\rm o}[i]{\boldsymbol v}^{*}[i],
\end{equation}
where the $M\times I$ Hankel matrix \cite{golub} with the received
samples of ${\boldsymbol r}[i]$ performs the convolution and is
described by
\begin{equation}
\boldsymbol{ \Re}_{\rm o}[i] = \left[\begin{array}{c c c c c}
r_{0}^{[i]} & r_{1}^{[i]}   & \ldots & r_{{\rm I}-1}^{[i]}  \\
r_{1}^{[i]}  & r_{2}^{[i]}   & \ldots & r_{{\rm I}}^{[i]}  \\
\vdots & \vdots  & \ddots & \vdots \\
r_{M-2}^{[i]}  & r_{M-1}^{[i]}  & \ldots & 0  \\
r_{M-1}^{[i]}  & 0  & \ldots & 0  \\
 \end{array}\right]. \label{hankel}
\end{equation}

The $M \times 1$ vector ${\boldsymbol r}_{\rm I}[i]$ is
transformed by a decimation unit that contains $B$ switching
decimation patterns in parallel, leading to $B$ different $D
\times 1$ vectors $\bar{\boldsymbol r}_{b}[i]$, $b=1, \ldots, B$,
where $L$ is the decimation factor and $D=M/L$ is the rank of the
BARC system. This is inspired by diversity techniques found in
wireless communications \cite{rappa}, whose principle is to
collect different copies of signals and combine them to increase
the signal-to-noise ratio, and switched control systems
\cite{liberzon} that exploit switching rules to stabilize and
design a system. The decimation procedure corresponds to
discarding $M-D$ samples of ${\boldsymbol r}_{\rm I}[i]$ with
different patterns, resulting in $B$ different $D \times 1$
decimated vectors $\bar{\boldsymbol r}_{b}[i]$. The $D \times 1$
decimated vector for branch $b$ is given by
\begin{equation}
\bar{\boldsymbol r}_{b}[i] = {\boldsymbol D}_{b}[i] {\boldsymbol
r}_{\rm I}[i],~~ b=1, \ldots, B
\end{equation}
where each row of ${\boldsymbol D}_{b}[i]$ contains a single $1$
and $M-1$ zeros. The $D\times M$ decimation matrix ${\boldsymbol
D}_{b}[i]$ is equivalent to removing $M-D$ samples of
${\boldsymbol r}_{\rm I}[i]$. The matrices ${\boldsymbol
D}_{b}[i]$ are designed off-line, stored at the receiver and the
best ${\boldsymbol D}_{b}[i]$ is selected to minimize a desired
objective function. The output $z_b[i]$ of the BARC scheme
corresponds to filtering $\bar{\boldsymbol r}_{b}[i]$ with
$\bar{\boldsymbol w}[i]$ and then selecting the branch that
minimizes the desired criterion. The output $z_b[i]$ is a function
of $\bar{\boldsymbol w}[i]$, ${\boldsymbol D}_b[i]$ and
${\boldsymbol v}[i]$ expressed by
\begin{equation}
\begin{split}
z_b[i] & = \bar{\boldsymbol w}^{H}[i]{\boldsymbol
S}_{D,b}^H[i]{\boldsymbol r}[i] =
\bar{\boldsymbol w}^{H}[i]\big({\boldsymbol D}_b[i]{\boldsymbol V}^{H}[i]{\boldsymbol r}[i]\big) \\
& = \bar{\boldsymbol w}^{H}[i] \big( {\boldsymbol
D}_b[i]\boldsymbol{\Re}_{\rm o}[i]\big) {\boldsymbol v}^{*}[i] =
\bar{\boldsymbol w}^{H}[i]
\boldsymbol{\Re}_b[i] {\boldsymbol v}^{*}[i]\\
& = {\boldsymbol
v}^{H}[i]\big(\boldsymbol{\Re}^T_b[i]\bar{\boldsymbol w}^{*}[i]
\big) = {\boldsymbol v}^{H}[i]{\boldsymbol u}[i],
\end{split}
\end{equation}
where ${\boldsymbol u}[i]=\boldsymbol{ \Re}^T_b[i]\bar{\boldsymbol
w}^{*}[i]$ is an $I\times 1$ vector, the $D$ coefficients of
$\bar{\boldsymbol w}[i]$ and the $I$ elements of ${\boldsymbol
v}[i]$ are assumed complex and the $D \times I$ matrix $
\boldsymbol{ \Re}_b[i]$ is $\boldsymbol{ \Re}_b[i] = {\boldsymbol
D}_b[i]\boldsymbol{\Re}_{\rm o}[i]$. In what follows, we will
develop constrained constant modulus (CCM)-based estimators and
describe how the switching rule is incorporated into the proposed
blind design.}

%The last processing stage of the proposed BARC scheme is the
%selection of the switching branch using as the metric the square
%of the constant modulus error for branch $b$ expressed by
%\begin{equation}
%e_{b}[i] = |\bar{\boldsymbol w}^{H}[i]{\boldsymbol
%S}_{D,b}^H[i]{\boldsymbol r}[i]|^2 -1 . \nonumber
%\end{equation}
%The selected decimation matrix ${\boldsymbol D}[i]$ is obtained as
%follows:
%\begin{equation}
%{\boldsymbol D}[i] = {\boldsymbol D}_{b_s}[i] ~~ \textrm{when} ~~
%b_s = \arg \min_{1\leq b \leq B} (e_{b}[i])^{2},
%\end{equation}
%where $B$ is the number of switching branches set in advance by
%the designer.

\subsection{ Joint Iterative CCM Design of Estimators and Discrete Optimization}

The design of the BARC scheme is equivalent to solving a joint
optimization problem with ${\boldsymbol v}[i]$, ${\boldsymbol
D}_b[i]$ and $\bar{\boldsymbol w}[i]$ using a strategy based on
fixing two parameters and optimizing one, and alternating the
procedure among the parameters until convergence. A key feature of
this problem is that it involves a combination of continuous and
discrete optimization procedures. Specifically, the design
corresponds to the constrained continuous minimization of the
estimators ${\boldsymbol v}[i]$ and $\bar{\boldsymbol w}[i]$ and
the discrete minimization of ${\boldsymbol D}[i]$ according to the
CCM design criterion.

{  Let us describe the CCM estimators design of the BARC
structure.} The CCM design for  ${\boldsymbol v}[i]$,
${\boldsymbol D}_b[i]$ and $\bar{\boldsymbol w}[i]$ can be
computed through the optimization problem
\begin{equation}
\begin{split}
 \big\{ {\boldsymbol v}_{\rm opt}, {\boldsymbol D}_{\rm opt}, \bar{\boldsymbol
w}_{\rm opt} \big\} & = \arg \min_{{\boldsymbol v}[i],{\boldsymbol
D}_b[i] ,\bar{\boldsymbol w}[i]} J_{\rm CM}({\boldsymbol
v}[i],{\boldsymbol D}_b[i] ,\bar{\boldsymbol
w}[i]) , \\
{\rm subject ~to~} & \bar{\boldsymbol w}_k^H[i] {\boldsymbol
S}_D^H[i]{\boldsymbol p}[i]= \nu, \label{optprob}
\end{split}
\end{equation}
where the parameter $\nu$ is a constant employed to enforce
convexity and
\begin{equation}
J_{\rm CM}({\boldsymbol v}[i],{\boldsymbol D}_b[i]
,\bar{\boldsymbol w}[i]) = E\Big[\big(|\bar{\boldsymbol
w}^{H}[i]\boldsymbol{\Re}[i]{\boldsymbol v}^{*}[i] |^{2} - 1
\big)^2\Big]. \label{barccost}
\end{equation}

The decimation matrix ${\boldsymbol D}_b[i]$ is selected to
minimize the square of the instantaneous constant modulus error
obtained for all the $B$ branches according to
\begin{equation}
{\boldsymbol D}_b[i] = {\boldsymbol D}_{b_{\rm s}}[i] ~~
\textrm{when} ~~ b_{\rm s} = \arg \min_{1\leq b \leq B}
(e_{b}[i])^{2}, \label{Ddesign}
\end{equation}
where the constant modulus error signal of the BARC scheme is $
e_{b}[i] = |\bar{\boldsymbol w}^{H}[i]{\boldsymbol
S}_{D,b}^H[i]{\boldsymbol r}[i]|^2 -1$. With the selected
decimation matrix ${\boldsymbol D}_b[i]$, we can form the
reduced-rank vector $\bar{\boldsymbol r}[i] = {\boldsymbol
D}_b[i]{\boldsymbol V}^{H}[i]{\boldsymbol r}[i]$ that will be used
in the following procedure for the design of ${\boldsymbol v}[i]$
and $\bar{\boldsymbol w}[i]$.

By using the method of Lagrange multipliers, fixing
$\bar{\boldsymbol w}[i]$ and minimizing the Lagrangian with
respect to ${\boldsymbol v}[i]$, the expression for the
interpolator becomes
\begin{equation}
\begin{split}
{\boldsymbol v}[i +1]  &= \bar{\boldsymbol R}^{-1}_u[i] \big[
\bar{\boldsymbol d}_u[i]
 - (\bar{\boldsymbol p}^H_w[i] \bar{\boldsymbol R}^{-1}_u[i] \bar{\boldsymbol p}_w[i])^{-1}  \\ & \quad \cdot
\bar{\boldsymbol p}_w[i] \big( \bar{\boldsymbol p}^H_w[i]
\bar{\boldsymbol R}^{-1}_u[i] \bar{\boldsymbol d}_u[i] - \nu \big)
\big], \label{vdesign}
\end{split}
\end{equation}
where $\bar{\boldsymbol R}_u[i] = E[|z[i]|^2 {\boldsymbol
u}[i]{\boldsymbol u}^{H}[i]]$, $\bar{\boldsymbol d}_u[i] =
E[z^*[i]{\boldsymbol u}[i]]$, ${\boldsymbol u}[i]=\boldsymbol{
\Re}^T_b[i]\bar{\boldsymbol w}^{*}[i]$ and $\bar{\boldsymbol
p}_w[i]= \boldsymbol{P}_o^T[i] \bar{\boldsymbol w}[i]$. {  The $D
\times {\rm I}$ matrix $\boldsymbol{P}_o[i]={\boldsymbol D}[i]
{\boldsymbol \Re}_p[i]$ arises from the constraint and the
equivalence $\bar{\boldsymbol w}_k^H[i] {\boldsymbol
S}_D^H[i]{\boldsymbol p}[i]= \bar{\boldsymbol w}_k^H[i]
\boldsymbol{P}_o^T[i] {\boldsymbol v}^*[i]= {\boldsymbol v}_k^H[i]
\bar{\boldsymbol p}_w[i]= \nu$, where ${\boldsymbol \Re}_p[i]$ is
a $D \times M$ Hankel matrix with elements of the effective
signature ${\boldsymbol p}[i]$ shifted in a similar way to
(\ref{hankel})}. By fixing the interpolator ${\boldsymbol v}[i]$
and minimizing the Lagrangian with respect to $\bar{\boldsymbol
w}[i]$, we obtain
\begin{equation}
\begin{split}
\bar{\boldsymbol w}[i+1]  &= \bar{\boldsymbol R}^{-1}_z[i] \big[
\bar{\boldsymbol d}_z[i]
 - (\bar{\boldsymbol p}^H[i] \bar{\boldsymbol R}^{-1}_z[i] \bar{\boldsymbol p}[i])^{-1}  \\ & \quad \cdot
\bar{\boldsymbol p}[i] \big( \bar{\boldsymbol p}^H[i]
\bar{\boldsymbol R}^{-1}_z[i] \bar{\boldsymbol d}_z[i] - \nu \big)
\big], \label{wdesign}
\end{split}
\end{equation}
where $\bar{\boldsymbol R}_z[i] = E[|z[i]|^2 \bar{\boldsymbol
r}[i]\bar{\boldsymbol r}^{H}[i]]={\boldsymbol
S}_D^H[i]{\boldsymbol R}_z[i]{\boldsymbol S}_D[i]$ , ${\boldsymbol
R}_z[i] = E[|z[i]|^2{\boldsymbol r}[i]{\boldsymbol r}^{H}[i]]$ ,
$\bar{\boldsymbol d}_z[i] = E[z^*[i]\bar{\boldsymbol r}[i] ] =
{\boldsymbol S}_D^H[i]E[z^*[i]{\boldsymbol r}[i]$,
$\bar{\boldsymbol p}[i]={\boldsymbol S}_D^H[i]{\boldsymbol p}[i]$
and ${\boldsymbol S}_D[i] = {\boldsymbol D}_b[i]{\boldsymbol
V}^H[i]$. We remark that (\ref{Ddesign}), (\ref{vdesign}) and
(\ref{wdesign}) depend on each other and their previous values.
Therefore, it is necessary to iterate (\ref{Ddesign}),
(\ref{vdesign}) and (\ref{wdesign}) in an alternated form (one
followed by the other) with an initial value to obtain a solution.
{  The expectations can be estimated either via time averages or
by instantaneous estimates as will be described by the adaptive
algorithms.}

\subsection{Design of Decimation Schemes }

%Having described the blind reduced-rank signal processing
%performed by the proposed BARC scheme, the design of the
%estimators and the selection of ${\boldsymbol D}_b[i]$, we are now
%interested in describing design strategies for the decimation
%matrix ${\boldsymbol D}_b[i]$. In the proposed BARC scheme, the
%basic idea is to employ $B$ decimation matrices ${\boldsymbol
%D}_b[i]$, $b=1,2, \ldots, B$ in parallel branches and select the
%branch that minimizes a design criterion. In this work, we focus
%on the constant modulus criterion, however, this can be extended
%to the mean squared error (MSE), minimum variance (MV) and minimum
%kurtosis criteria.

We are interested in developing decimation schemes that are
cost-effective and easy to employ with the proposed BARC scheme.
This can be done by imposing constraints on the structure of
${\boldsymbol D}_b[i]$. Since the operator ${\boldsymbol D}_b[i]$
performs decimation, the structure of ${\boldsymbol D}_b[i]$ is
constrained to contain only zeros and $D$ ones. Thus, the
decimation operation of the BARC scheme amounts to discarding
samples in conjunction with filtering by ${\boldsymbol v}[i]$ and
$\bar{\boldsymbol w}[i]$. The decimation matrix ${\boldsymbol
D}_b[i]$ is selected so to minimize the square of the
instantaneous constant modulus error obtained for the $B$ branches
employed as follows
\begin{equation}
{\boldsymbol D}_b[i] = {\boldsymbol D}_{b_{\rm s}}[i] ~~
\textrm{when} ~~ b_{\rm s} = \arg \min_{1\leq b \leq B}
(e_{b}[i])^{2}, \label{Ddesign2}
\end{equation}
where $ e_{b}[i] = |\bar{\boldsymbol w}^{H}[i]{\boldsymbol
S}_{D,b}^H[i]{\boldsymbol r}[i]|^2 -1$. The design of the
decimation matrix ${\boldsymbol D}_b[i]$ considers a general
framework that can be used for any decimation scheme and is
illustrated by
\begin{equation}
{\boldsymbol D}_{b}[i] = \left[\hspace*{0.5em}\begin{array}{c}
{\boldsymbol d}^T_{1,b}[i]  \\
\vdots   \\
{\boldsymbol d}^T_{j,b}[i]  \\
\vdots  \\
{\boldsymbol d}^T_{D,b}[i]
\end{array}\hspace*{0.5em}\right], \label{decimat}
\end{equation}
where each row of the matrix ${\boldsymbol D}_b[i]$ is structured
as
\begin{equation}
{\boldsymbol d}_{j,b}[i] = [\underbrace{0~~ \ldots ~~
0}_{\gamma_{j}~zeros} ~~~ 1 ~~  \underbrace{ 0  ~~~ \ldots ~~
0}_{(M-\gamma_{j}-1) ~zeros} ]^T, \label{decvec}
\end{equation}
and the index $j$ ($j=1,2,\ldots, D$) denotes the $j$-th row of
the matrix, the rank of the matrix ${\boldsymbol D}_b[i]$ is
$D=M/L$, the decimation factor is $L$ and $B$ corresponds to the
number of parallel branches. The quantity $\gamma_{j}$ is the
number of zeros chosen according to a given design criterion.

Given the constrained structure of ${\boldsymbol D}_b[i]$, it is
possible to devise an optimal procedure for designing
${\boldsymbol D}_b[i]$ via an exhaustive search of all possible
design patterns with the adjustment of the variable $\gamma_{j}$,
where an exhaustive procedure that selects $D$ samples out of $M$
possible candidates is performed. The total number of patterns
$B_{\rm ex}$ is equal to
\begin{equation}
{ B}_{\rm ex} = \underbrace{M\cdot(M-1) \ldots (M-D+1)}_{D ~~
\textrm{terms}} = \left( \begin{array}{c} M
\\ D \end{array} \right). \nonumber
\end{equation}
We can view this exhaustive procedure as a combinatorial problem
that has $M$ samples as possible candidates for the first row of
${\boldsymbol D}_b[i]$ and considers $M-j+1$ positions as
candidates for the following $D-1$ rows of the matrix
${\boldsymbol D}_b[i]$, where $j$ is the index used to denote
$j$th row of the matrix ${\boldsymbol D}_b[i]$. The exhaustive
scheme described above is, however, too complex for practical use
because it requires $D$ permutations of $M$ samples for each
symbol interval and $M-1$ candidates for the positions, and
carries out an extensive search over all possible patterns.

It is highly desirable to employ decimation schemes that are
cost-effective and gather important properties such as
low-requirements of storage and computational complexity and can
work with a small number of branches $B$. By adjusting the
variable $\gamma_{j}$ in the framework depicted in
(\ref{decimat}), we can obtain the following sub-optimal schemes:

\begin{description}
\item[${\boldsymbol A.}$]{Uniform (U) Decimation with $B=1$. We make $\gamma_{j} =
(j-1)L$ and this corresponds to the use of a single branch ($B=1$)
on the decimation unit (no switching and optimization of
branches), and is equivalent to the scheme in \cite{delamaretvt}.}

\item[${\boldsymbol B.}$]{Pre-Stored (PS) Decimation. We select $\gamma_{j} =
(j-1)L+(b-1)$ which corresponds to the utilization of uniform
decimation for each branch $b$ out of $B$ branches and the
different patterns are obtained by picking out adjacent samples
with respect to the previous and succeeding decimation patterns.}

\item[${\boldsymbol C.}$]{Random (R) Decimation. We choose $\gamma_{j}$ as a
discrete uniform random variable, which is independent for each
row $j$ out of $B$ branches and whose values range between $0$ and
$M-1$. {  A constraint is included to avoid rows with repetitive
patterns.} }
\end{description}

\section{Blind Adaptive Estimation Algorithms}

{  In this section, we develop SG and RLS estimation algorithms
\cite{haykin} for estimating the parameters of the BARC scheme
(${\boldsymbol v}[i]$,${\boldsymbol D}[i]$ and $\bar{\boldsymbol
w}[i]$). The SG algorithms require the setting of step sizes and
are indicated for situations where the eigenvalue spread of
$\bar{\boldsymbol R}_z^{-1}[i]$ is small. The RLS algorithms need
the setting of forgetting factors and are suitable for scenarios
in which $\bar{\boldsymbol R}_z^{-1}[i]$ has a large eigenvalue
spread. We also present blind model-order selection algorithms for
adjusting the lengths $D$ and $I$ of the estimators and algorithms
for determining the minimum number of branches required to achieve
a predetermined performance. The model-order and number of
branches selection algorithms are decoupled in order to reduce the
search space and the computational cost. We have tested a joint
search over $I$, $D$ and $B$ and this has not resulted in
performance gains over the separate search over $B$ and over $I$
and $D$.  Unlike prior work \cite{jidf} with the MSE criterion,
the proposed algorithms employ the CM approach and rely on a set
of linear constraints. The complexity of the proposed SG, RLS and
model-order selection algorithms is compared with existing methods
in terms of additions and multiplications. }

\subsection{SG Algorithms for The BARC Scheme}

To design the estimators ${\boldsymbol v}[i]$ and
$\bar{\boldsymbol w}[i]$ and the decimation matrix ${\boldsymbol
D}[i]$, we consider the Lagrangian
\begin{equation}
\begin{split}
{\mathcal{L}}({\boldsymbol v}[i],{\boldsymbol
D}[i],\bar{\boldsymbol w}[i]) & = E\Big[\big(|\bar{\boldsymbol
w}^{H}[i]\boldsymbol{\Re}_b[i]{\boldsymbol v}^{*}[i] |^{2} - 1
\big)\Big] \\ & \quad  + 2\Re ~ \Big[\big(\bar{\boldsymbol w}^H[i]
{\boldsymbol S}_D^H[i]{\boldsymbol p}[i] - \nu \big) \lambda
\Big], \label{lagbarc}
\end{split}
\end{equation}
where $\lambda$ is a Lagrange multiplier and ${ \Re}[ \cdot]$
denotes the real part of the argument. The input vector
${\boldsymbol r}[i]$ is processed by the interpolator
${\boldsymbol v}[i]$, yielding ${\boldsymbol r}_{\rm I}[i] =
{\boldsymbol V}^H[i]{\boldsymbol r}[i]$. We then compute the
decimated interpolated vectors ${\boldsymbol r}_{b}[i]$ for the
$B$ branches with the decimation matrix ${\boldsymbol D}_{b}[i]$,
where $1 \leq b \leq B$. Once the $B$ candidate vectors
$\bar{\boldsymbol r}_{b}[i]$ are computed, we select the vector
$\bar{\boldsymbol r}_{b}[i]$ which minimizes the square of
\begin{equation}
e_{b}[i]=|\bar{\boldsymbol w}^{H}[i]{\boldsymbol
S}_{D,b}[i]{\boldsymbol r}[i]|^2 -1.
\end{equation}
where ${\boldsymbol S}_{D,b}[i] = {\boldsymbol V}[i] {\boldsymbol
D}_b^H[i]$. Based on the selection of ${\boldsymbol D}_b[i]$, we
choose the corresponding reduced-rank vector $\bar{\boldsymbol
r}[i]$ and select the error of the proposed SG algorithm $e[i]$ as
the error $e_{b}[i]$ with the smallest squared magnitude of the
$B$ branches according to
\begin{equation}
\begin{split}
{\boldsymbol S}_{D}[i] & = {\boldsymbol S}_{D,b_{s}},~~
\bar{\boldsymbol r}[i] = \bar{\boldsymbol
r}_{b_s}[i] ~ {\rm and} ~ e[i] = e_{b_s}[i] \\ & ~~~ {\rm when} \\
b_s & = \arg \min_{1 \leq b \leq B} (e_{b}[i])^{2}. \label{sgsel}
\end{split}
\end{equation}
{  In order to derive an SG algorithm for ${\boldsymbol v}[i]$, we
need to transform the proposed constraint in (\ref{optprob}) and
obtain a suitable and equivalent form for use with ${\boldsymbol
v}[i]$. We can write $\bar{\boldsymbol w}^H[i] {\boldsymbol
S}_D^H[i]{\boldsymbol p}[i]= \bar{\boldsymbol w}^H[i]
\boldsymbol{P}_o^T[i] {\boldsymbol v}^*[i]= {\boldsymbol v}^H[i]
\bar{\boldsymbol p}_w[i]= \nu$, where $\bar{\boldsymbol p}_w[i]=
\boldsymbol{P}_o^T[i] \bar{\boldsymbol w}[i]$ and the $D \times
{\rm I}$ matrix $\boldsymbol{P}_o[i]$ is a function of
${\boldsymbol D}_b[i]$ and ${\boldsymbol p}[i]$ and is given by
$\boldsymbol{P}_o[i] = {\boldsymbol D}[i] {\boldsymbol \Re}_p[i]$,
where ${\boldsymbol \Re}_p[i]$ is a $D \times M$ Hankel matrix
with elements of the effective signature ${\boldsymbol p}[i]$
shifted in a similar way to (\ref{hankel}). We need to construct
$\bar{\boldsymbol p}_w[i]$ for each symbol from ${\boldsymbol
P}_o[i]$ and $\bar{\boldsymbol w}[i]$. Minimizing (\ref{lagbarc})
and using the proposed equivalent constraint ${\boldsymbol v}^H[i]
\bar{\boldsymbol p}_w[i]= \nu$, we obtain}
\begin{equation}
{\boldsymbol v}[i+1] = {\boldsymbol v}[i] - \mu_v e[i] z^*[i]
\bigg( {\boldsymbol I} - (\bar{\boldsymbol
p}_w^H[i]\bar{\boldsymbol p}_w[i])^{-1} \bar{\boldsymbol
p}_w[i]\bar{\boldsymbol p}_w^H[i] \bigg) {\boldsymbol u}[i],
\label{sgrec}
\end{equation}
where $\mu_v$ is the step size. Minimizing (\ref{lagbarc}) and
using the constraint $\bar{\boldsymbol w}^H[i] {\boldsymbol
S}_D^H[i]{\boldsymbol p}[i]= \nu$, we obtain
\begin{equation}
\begin{split}
\bar{\boldsymbol w}[i+1] & = \bar{\boldsymbol w}[i] - \mu_w e[i]
z^*[i] \bigg( {\boldsymbol I} - (\bar{\boldsymbol
p}^H[i]\bar{\boldsymbol p}[i])^{-1}  \bar{\boldsymbol
p}[i]\bar{\boldsymbol p}^H[i]\bigg) \bar{\boldsymbol r}[i],
\label{wrecsgj}
\end{split}
\end{equation}
where $\mu_w$ is the step size. {  The SG algorithm for the BARC
has a computational complexity $O(D + N_{I})$ and employs
equations (\ref{sgsel})-(\ref{wrecsgj}). In fact, the BARC scheme
trades off one SG algorithm with complexity $O(M)$ against two SG
algorithms with complexity $O(D)$ and $O({\rm I})$, operating
simultaneously and exchanging information.}

%${\boldsymbol \Re}_p[i] = \left[\begin{array}{cccc}
%  p_0[i] & p_1[i] & \ldots & p_{{\rm I}-1}[i] \\
%  \vdots & \ddots & \ddots & \vdots \\
%  p_{M-2} & p_{M-1}[i] & \ldots & 0\\
%  p_{M-1} & 0 & \ldots & 0[i]
%\end{array}\right].$

\subsection{RLS Algorithms for the BARC Scheme}

In order to design the estimators ${\boldsymbol v}[i]$,
$\bar{\boldsymbol w}[i]$ and the matrix ${\boldsymbol D}[i]$ with
RLS algorithms, we consider the Lagrangian
\begin{equation}
\begin{split}
{\mathcal{L}}_{\rm LS}({\boldsymbol v}[i],{\boldsymbol D}[i],
\bar{\boldsymbol w}[i]) & = \sum_{l=1}^{i} \alpha^{i-l}
\big(|\bar{\boldsymbol w}^{H}[i]\boldsymbol{\Re}[l]{\boldsymbol
v}^{*}[i] |^{2} - 1 \big) \\ & \quad  + 2\Re
~\Big[\big(\bar{\boldsymbol w}^H[i] {\boldsymbol
S}_D^H[i]{\boldsymbol p}[i] - \nu \big) \lambda \Big],
\label{lagbarc2}
\end{split}
\end{equation}
where $\lambda$ is a Lagrange multiplier and $\alpha$ is a
forgetting factor. We perform the signal processing according to
the block diagram of Fig. 1. Based on the choice of ${\boldsymbol
D}_b[i]$, we select the corresponding reduced-rank vector
$\bar{\boldsymbol r}[i]$ and the error $e[i]$ as the error
$e_{b}[i]=|\bar{\boldsymbol w}^{H}[i]{\boldsymbol
S}_{D,b}[i]{\boldsymbol r}[i]|^2 -1$ with the smallest squared
magnitude of the $B$ branches as follows
\begin{equation}
\begin{split}
{\boldsymbol S}_{D}[i] & = {\boldsymbol S}_{D,b_{s}},~~
\bar{\boldsymbol r}[i] = \bar{\boldsymbol
r}_{b_s}[i] ~ {\rm and} ~ e[i] = e_{b_s}[i] \\ & ~~~ {\rm when} \\
b_s & = \arg \min_{1 \leq b \leq B} (e_{b}[i])^{2}. \label{rlssel}
\end{split}
\end{equation}
Minimizing (\ref{lagbarc2}) with respect to ${\boldsymbol v}[i]$,
using the constraint ${\boldsymbol v}^H[i] {\boldsymbol p}_w[i]=
\nu$ and the matrix inversion lemma \cite{haykin}, we get
\begin{equation}
\begin{split}
{\boldsymbol v}[i+1] & = \hat{\boldsymbol R}_{u}^{-1}[i]
\bigg(\hat{\boldsymbol d}_{u}[i] + ({\boldsymbol p}_w^H[i]
\hat{\boldsymbol R}_{u}^{-1}[i] {\boldsymbol p}_{w}[i])^{-1}  \\
& \quad \cdot {\boldsymbol p}_w[i] ({\boldsymbol
d}_w^H[i]\hat{\boldsymbol R}_{u}^{-1}[i] {\boldsymbol p}_{w}[i] -
\nu) \bigg),
\end{split}
\end{equation}
where
\begin{equation}
\begin{split}
\hat{\boldsymbol d}_u[i] = \alpha \hat{\boldsymbol d}_u[i-1] +
(1-\alpha) z^*[i] {\boldsymbol u}[i]
\end{split}
\end{equation}
\begin{equation}
{\boldsymbol k}_{u}[i] = \frac{\alpha^{-1} \hat{\boldsymbol
R}_{u}^{-1} [i-1] z[i] {\boldsymbol u}[i]}{1+\alpha^{-1}
{\boldsymbol u}^H[i]z[i] \hat{\boldsymbol
R}_{u}^{-1}[i-1]z^*[i]{\boldsymbol u}[i]}
\end{equation}
\begin{equation}
\hat{\boldsymbol R}_{u}^{-1}[i] = \alpha^{-1} \hat{\boldsymbol
R}_{u}^{-1}[i-1] - \alpha^{-1} {\boldsymbol k}_{u}[i]
z^*[i]{\boldsymbol u}^H[i] \hat{\boldsymbol R}^{-1}_u[i-1]
\label{Ru}
\end{equation}
and the initial values of the recursions are $\hat{\boldsymbol
R}_{u}^{-1}[i] = \delta_v {\boldsymbol I}$ and $ \hat{\boldsymbol
d}_u[0] = \rho_v$, where $\delta_v$ and $\rho_v$ are small
positive scalars. Minimizing (\ref{lagbarc2}) with respect to
$\bar{\boldsymbol w}[i]$, using the constraint $\bar{\boldsymbol
w}_k^H[i] {\boldsymbol S}_D^H[i]{\boldsymbol p}[i]= \nu$ and the
the matrix inversion lemma \cite{haykin}, we obtain
\begin{equation}
\begin{split}
{\bar{\boldsymbol w}}[i+1] & = {\hat{\bar{\boldsymbol
R}}}_z^{-1}[i] \bigg( {\hat{\bar{\boldsymbol d}}}_z[i] +
({\bar{\boldsymbol p}}^H[i]{\hat{\bar{\boldsymbol
R}}}_z^{-1}[i]{{\bar{\boldsymbol p}}}[i])^{-1} \\ & \quad \cdot
{\bar{\boldsymbol p}}[i] ({\bar{\boldsymbol
p}}^H[i]{\hat{\bar{\boldsymbol
R}}}_z^{-1}[i]{\hat{\bar{\boldsymbol d}}}_z[i] - \nu) \bigg),
\label{wrecrlsj}
\end{split}
\end{equation}
where
\begin{equation}
\begin{split}
{\hat{\bar{\boldsymbol d}}}_z[i] = \alpha {\hat{\bar{\boldsymbol
d}}}_z[i-1] + (1-\alpha) z^*[i] \bar{\boldsymbol r}[i]
\end{split}
\end{equation}
\begin{equation}
\bar{\boldsymbol k}_{z}[i] = \frac{\alpha^{-1}
{\hat{\bar{\boldsymbol R}}}_z^{-1} [i-1] z[i] \bar{\boldsymbol
r}[i]}{1+\alpha^{-1} \bar{\boldsymbol r}^H[i]z[i]
{\hat{\bar{\boldsymbol R}}}_z^{-1}[i-1]z^*[i]\bar{\boldsymbol
r}[i]}
\end{equation}
\begin{equation}
{\hat{\bar{\boldsymbol R}}}_z^{-1}[i] = \alpha^{-1}
{\hat{\bar{\boldsymbol R}}}_z^{-1}[i-1] - \alpha^{-1}
\bar{\boldsymbol k}_{z}[i] z^*[i]\bar{\boldsymbol r}^H[i]
{\hat{\bar{\boldsymbol R}}}_z^{-1}[i-1] \label{Rz}
\end{equation}
and the initial values of the recursions are $\hat{\boldsymbol
R}_{z}^{-1}[i] = \delta_w {\boldsymbol I}$ and $ \hat{\boldsymbol
d}_z[0] = \rho_w$, where $\delta_w$ and $\rho_w$ are small
positive scalars. {  The RLS algorithm for the BARC has a
computational cost of $O(D^2) + O({ I}^2)$ and
consists of equations (\ref{rlssel})-(\ref{Rz}).} %In fact, the BARC
%scheme trades off one RLS algorithm with complexity $O(M^2)$
%against two RLS algorithms with complexity $O(D^2)$ and $O({
%I}^2)$, operating simultaneously and exchanging information.

\subsection{Model-Order Selection Algorithms}

{  This part develops model-order selection algorithms for
automatically adjusting the lengths of the estimators used in the
BARC scheme. Prior work in this area has focused on methods for
model-order selection which utilize MSWF-based algorithms
\cite{goldstein} or AVF-based recursions \cite{avf5,avf6,avf7}. In
the proposed approach we constrain the search within a range of
appropriate values and rely on a CCM-based LS criterion to
determine the lengths of ${\boldsymbol v}[i]$ and
$\bar{\boldsymbol w}[i]$ that can be adjusted in a flexible
structure. The proposed scheme with extended filters is
significantly less complex than the multiple filters approach
reported in \cite{jidf}.} The model-order selection algorithm for
the BARC is called Auto-Rank and minimizes
\begin{equation} {\mathcal
C}({\boldsymbol v}[i], {\boldsymbol D}[i],\bar{\boldsymbol w}[i])
= \sum_{l=1}^{i} \alpha^{i-l} \Big( |\bar{\boldsymbol
w}^{H}[i]{\boldsymbol D}[i]\boldsymbol{\Re}_o[l]{\boldsymbol
v}^{*}[i] |^{2} - 1 \Big) , \label{eq:costadap2}
\end{equation}
The order of ${\boldsymbol v}[i]$, ${\boldsymbol D}[i]$,
$\bar{\boldsymbol w}[i]$, and the associated matrices
$\hat{\bar{\boldsymbol R}}_u[i]$, and $\hat{\bar{\boldsymbol
R}}_z[i]$ defined in (\ref{Ru}) and (\ref{Rz}), respectively, that
are necessary for the computation of ${\boldsymbol v}[i]$ and
$\bar{\boldsymbol w}[i]$ require adjustment. To this end, we
predefine ${\boldsymbol v}[i]$ and $\bar{\boldsymbol w}[i]$ as
follows:
\begin{equation}
\begin{split}
{\boldsymbol v}[i] & = \left[\begin{array}{cccccc} v_1[i] &
v_2[i] & \ldots & v_{{ I_{\rm min}}}[i] & \ldots & v_{ I_{{\rm max}}}[i] \end{array}\right]^T \\
\bar{\boldsymbol w}[i] & = \left[\begin{array}{cccccc} w_1[i] &
w_2[i] & \ldots & w_{D_{\rm min}}[i] & \ldots & w_{D_{\rm max}}[i]
\end{array}\right]^T
\end{split}
\end{equation}

For each data symbol we select the best order for the model. The
proposed Auto-Rank algorithm that chooses the best lengths ${
D}_{\rm opt}[i]$ and ${I_{\rm opt}}[i]$ for the filters
${\boldsymbol v}[i]$ and $\bar{\boldsymbol w}[i]$, respectively,
is given by
\begin{equation}
\{ D_{\rm opt}[i], {I_{\rm opt}}[i] \} = \arg
\min_{\underset{D_{\rm min} \leq d \leq D_{\rm max}}{{I_{\rm min}}
\leq n \leq {I_{\rm max}}}} {\mathcal C}({\boldsymbol v}[i],
{\boldsymbol D}[i],\bar{\boldsymbol w}[i])
\end{equation}
where $d$ and $n$ are integers, $D_{\rm min}$ and $D_{\rm max}$,
and ${I_{\rm min}}$ and ${I_{\rm max}}$ are the minimum and
maximum ranks allowed for the reduced-rank filter and the
interpolator, respectively. %Note that a smaller rank may provide
%faster adaptation during the initial stages of the estimation
%procedure and a slightly greater rank usually yields a better
%steady-state performance. Our studies reveal that the range for
%which the ranks $D$ and $I$ of the proposed algorithms have a
%positive impact on the performance of the algorithms are limited,
%being from ${I_{\rm min}}=2$ to ${I_{\rm max}}=6$ for the
%interpolator and from $D_{\rm min}=3$ to $D_{\rm max}=6$ for the
%reduced-rank filter recursions. These values are rather
%insensitive to the system load (number of users), to the
%processing gain and work very well for all scenarios examined.
The additional complexity of the Auto-Rank algorithm is that it
requires the update of all involved quantities with the maximum
allowed rank $D_{\rm max}$ and ${I_{\rm max}}$ and the computation
of the cost function in (\ref{eq:costadap2}). This procedure can
significantly improve the convergence performance and can be
relaxed (the rank can be made fixed) once the algorithm reaches
steady state. An inadequate rank for adaptation may lead to a
performance degradation, which gradually increases as the
adaptation rank
deviates from the optimal rank. %In the simulations section, we
%will illustrate how the proposed rank adaptation algorithm
%performs.

\subsection{Automatic Selection of the Number of Branches}

{  In this subsection we propose algorithms for automatically
selecting the number of branches necessary to achieve a
predetermined performance. This performance measure is determined
off-line as a quantity related to the constant modulus cost
function. The first algorithm, termed selection of the number of
branches (SNB), relies on a simple search over the parallel
branches of the BARC scheme and tests whether the predetermined
performance has been attained via a comparison with a threshold
$\rho$. The second algorithm builds on the SNB algorithm and
incorporates prior statistical knowledge about the use of the
branches via sorting and is denoted SNB-S. Let us first define for
each time interval $i$ the branch cost as}
\begin{equation}
{\mathcal C}_{\rm branch}({\boldsymbol v}[i],~ {\bf D}_b[i],
\bar{\boldsymbol w}[i] ) = (e_b[i])^2 \label{eq:costinst}
\end{equation}
where $$e_b[i] = |\bar{\boldsymbol w}^{H}[i] {\boldsymbol
D}_b[i]\boldsymbol{\Re}_{\rm o}[i] {\boldsymbol v}^{*}[i] |^{2} -
1 $$ is the error signal for each branch. The proposed algorithms
for automatically selecting the number of branches perform the
following optimization
\begin{equation}
\begin{split}
 B_{\rm s}[i]  = \arg \min_{B_{\rm max}} ~~ \min_{1 \leq b \leq B_{\rm max}}
 ~
{\mathcal C}_{\rm branch}({\boldsymbol v}[i],~ {\bf D}_b[i], \bar{\boldsymbol w}[i]) \\
{\rm subject ~~to}~ {\mathcal C}_{\rm branch}({\boldsymbol v}[i],~
{\bf D}_b[i], \bar{\boldsymbol w}[i]) \leq \rho \label{eq:branch}
\end{split}
\end{equation}
where $b$ is an integer and $B_{\rm max}$ is the maximum number of
branches allowed for the BARC scheme, respectively, $B_{\rm s}$ is
the number of branches required to attain the desired performance
and $\rho$ is the prespecified performance. The SNB algorithm
determines the minimum number of branches necessary to achieve a
predetermined performance $\epsilon$ according to the cost
function defined in (\ref{eq:costinst}). It iteratively increases
the number of branches by one until the predetermined performance
$\rho$ is attained. The parameter $\rho$ can be chosen as a
function of the MMSE with a penalty allowed by the designer. An
alternative to the SNB algorithm is to exploit prior statistical
knowledge about the most frequently used branches and sort the
decimation matrices ${\boldsymbol D}_b[i]$ in descending order of
probability of occurrence. The SNB algorithm with sorting will be
termed SNB-S and consists of ordering the matrices ${\boldsymbol
D}_b[i]$ which are most likely to be used. This can be done at the
beginning of the transmission
and updated whenever required. %A summary of the proposed algorithms
%for selecting the minimum number of branches is shown in Table
%III.
An important measure that arises from the SNB and SNB-S
algorithms is the average number of branches $B_{\rm{avg}} = 1/Q
\sum_{i=1}^Q B_s[i]$ with $Q$ being the data record, which
illustrates the savings in computations of the branches.

\subsection{Computational Complexity}

{  In this section we detail the computational complexity of the
proposed and existing SG, RLS and model-order selection
algorithms, as shown in Tables I, II and III. This complexity
refers to an adaptive linear receiver that only requires the
timing and the spreading code of the user of interest. The
computational requirements are described in terms of additions and
multiplications and have been derived by counting the necessary
operations to compute each of the recursions required by the
analyzed algorithms. The key parameters of the complexity are the
length $D$ of $\bar{\boldsymbol w}[i]$ or the number of auxiliary
vectors (AVs) for the AVF algorithm \cite{avf5,avf6,avf7}, the
number of samples $M$ of ${\boldsymbol r}[i]$, the number of
branches $B$, the length $I$ of ${\boldsymbol v}[i]$ and the
number $L_p$ of assumed multipath components.

\begin{table}[h]
\centering%
\caption{\normalsize Computational complexity of SG algorithms. }
{
\begin{tabular}{ccc}
\hline \rule{0cm}{2.5ex}&  \multicolumn{2}{c}{Number of operations
per symbol } \\ \cline{2-3}
Algorithm & Additions & Multiplications \\
\hline
\emph{\small \bf Full-rank-trained \cite{haykin}} &  {\small $2M$} & {\small $2M+1$}  \\
\emph{\small  (eq. (9.5)-(9.7) of \cite{haykin})} &  {\small $2M$} & {\small $2M+1$}  \\
\emph{\small \bf MSWF-trained \cite{goldstein}} & {\small $2(D-1)^2 + 2D(M-1) $} & {\small $D^2 + 3D + 2DM $} \\
\emph{\small (eq. (53)-(62) of \cite{goldstein})} & {\small $ + (D-1)(M-1) + M$} & {\small $+ M + 1$} \\
\emph{\small \bf Full-rank-CCM \cite{xu&liu}} & {\small $8M + ML_{p} $} & {\small $7M +ML_{p} $} \\
\emph{\small (eq. (10),(11),(13) of \cite{xu&liu})} & {\small $+2$} & {\small $+2$} \\
\emph{\small \bf MSWF-CCM \cite{mswfccm}} & {\small $DM^2+3(D-1)^2+ 2D $} & {\small $DM^2+ 2D^2 + 7D  $}  \\
\emph{\small (table II of \cite{mswfccm})} & {\small $+ 3DM  +4M+3$} & {\small $+ 2DM+ ML_p +2$}  \\
\emph{\small \bf JIO-CCM \cite{jio}} & {\small $4DM + M $} & {\small $4DM + M  $}  \\
\emph{\small (eq. (14)-(15) of \cite{jio})} & {\small $+ 2D - 2$} & {\small $+ 7D + 6$}  \\
\emph{\small \bf Proposed BARC-CCM} & {\small $4D+ BD +4I $} & {\small $4D + B(D+1) $}  \\
\emph{\small (eq. (18)-(21)} & {\small $+ (I-1)M-2$} & {\small $+ 5I+IM + 4$}  \\
\hline

\end{tabular}
}
\end{table}

\begin{table}[h]
\centering%
\caption{\normalsize Computational complexity of RLS and AVF-based
algorithms. } {
\begin{tabular}{ccc}
\hline \rule{0cm}{2.5ex}&  \multicolumn{2}{c}{Number of operations
per symbol } \\ \cline{2-3}
Algorithm & Additions & Multiplications \\
\hline
\emph{\small \bf Full-rank-trained \cite{haykin}} & {\small $3(M-1)^{2} $} & {\small $3M^{2}$}   \\
\emph{\small (table 13.1 of \cite{haykin}) } & {\small $ + M^{2} + 2M$} & {\small $+2M + 2$}   \\
\emph{\small \bf MSWF-trained \cite{goldstein}}  & {\small $D^2 + 2(D-1)^2 $} & {\small $4D^2 + 3D  $} \\
\emph{\small (eq. (69)-(71) of \cite{goldstein})}  & {\small $+ 2D(M-1)+ M $} & {\small $+ 2DM+ ML_p$} \\
\emph{\small }  & {\small $+(D-1)(M-1) $} & {\small $+4$} \\
\emph{\small \bf Full-rank-CCM \cite{delamareccm}}  & {\small $5(M-1)^{2}$} & {\small $4M^{2}+5M$} \\
\emph{\small (eq. (6)-(10) of \cite{delamareccm})}  & {\small $+M^{2}+5M-1$} & {\small $+L_{p}^{2}+ML_{p}$} \\
\emph{\small }  & {\small $+3(L_{p}-1)^{2}$} & {\small $+L_{p}+4$} \\
\emph{\small \bf MSWF-CCM \cite{mswfccm}}  & {\small $DM^2 + 5D^2  $} & {\small $DM^2 + 6D^2  $} \\
\emph{\small  (table III of \cite{mswfccm}) } & {\small $+ 2DM + (M-1)L_p $} & {\small $+ 2DM +(D-1)M $} \\
\emph{} & {\small $+ (D-1)(M+1) $} & {\small $ + 7D  + ML_p+4$} \\
\emph{\small \bf JIO-CCM \cite{jio}} & {\small $5M^2 + DM$} & {\small $6M^2 + (2D + 6)M $}  \\
\emph{\small (eq. (10)-(11) of \cite{jio})} & {\small $+ 5D^2 + 3D - 1$} & {\small $+5D^2 + 9D + 3$}  \\
\emph{\small \bf AVF-trained \cite{avf5}} & {\small $D(3M^2-2M)$} & {\small $D(4M^2+3M)$} \\
\emph{\small (eq. (3),(11)-(13) of \cite{avf5}} & {\small $ +2M-1$} & {\small $+4M+2$} \\
\emph{\small \bf Proposed BARC-CCM} & {\small $6D^2 +(B+1) D +6I^2  $} & {\small $7D^2 + (B+8)D + 7I^2 $}  \\
\emph{\small (eq. (23)-(31))} & {\small $+I+ (I-1)M +8$} & {\small $+7I+ IM +3$}  \\
%\emph{\small \bf AVF-CMV \cite{avf5}} & {\small $D(M^2+3(M-1)^2) $} & {\small $D(4M^2+4M + 1)$} \\
%\emph{\small } & {\small $ +D(5(M-1)+1)$} & {\small $+ ML_p+ 2M + 1$} \\
%\emph{\small } & {\small $ +(M-1) +M(L_p-1)$} & {\small $$} \\
\hline
\end{tabular}
}
\end{table}

{  In Fig. 2 we illustrate the main complexity trends by showing
the computational complexity in terms of the arithmetic operations
as a function of the number of samples $M$. We use the same colors
for the corresponding SG techniques in Fig. 2 (a) and the RLS
counterparts in Fig. 2 (b). For these curves, we consider
$L_{p}=9$, $D=5$, $I=3$ and $B=8$ for the BARC, assume $D=4$ for
the MSWF-SG based approaches, while we use $D=5$ for the MSWF-RLS
techniques and $D=8$ for the AVF technique with non-orthogonal
auxiliary vectors (AVs) \cite{avf5,avf6,avf6}. The reason why we
use different values for $D$ is because we must find the most
appropriate trade-off between the model bias and variance
\cite{scharf} by adjusting $D$ (AVs for the AVF) and this depends
on the scheme.} We always use the best values for each scheme. The
curves in Fig. 2 (a) show that the reduced-rank BARC SG algorithms
have a complexity slightly higher than the full-rank trained SG
algorithms and substantially lower than the other analyzed
reduced-rank algorithms. For the RLS algorithms, depicted in Fig.
2 (b), we verify that the BARC reduced-rank scheme is much simpler
than any full-rank or reduced-rank RLS algorithm. This is because
there is a quadratic cost on $M$ rather than $D$ for the full-rank
schemes operating with the RLS algorithm and a high computational
cost associated with the design of the transformation matrix
${\boldsymbol S}_D[i]$ for all reduced-rank methods except for the
BARC scheme. The AVF scheme \cite{avf5,avf6,avf7} usually requires
extra complexity as it has more operations per auxiliary vector
(AV) and also requires a higher number of AVs to ensure a good
performance. The trained AVF employs a cross-correlation vector
estimated by $\hat{\bf p}[i] = \alpha \hat{\bf p}[i-1] +
(1-\alpha) b_k^*[i]{\bf r}[i]$.

\begin{figure}[!htb]
\begin{center}
\def\epsfsize#1#2{1\columnwidth}
\epsfbox{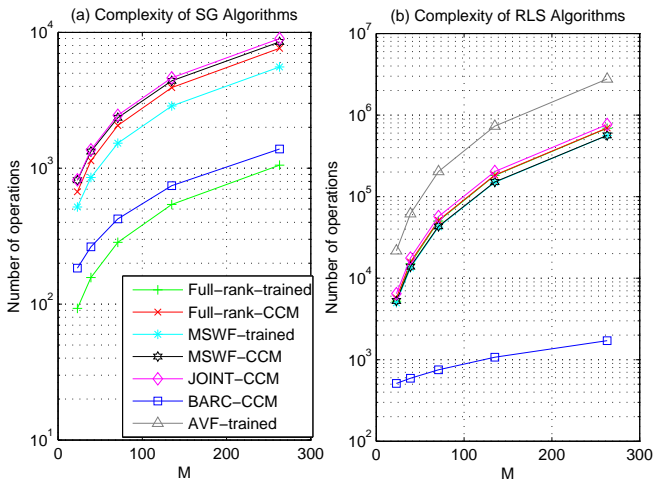} \vspace{-0.75em}\caption{Complexity in terms of
arithmetic operations of (a) SG and (b) RLS algorithms and AVF-based
recursions.}
\end{center}
\end{figure}

\begin{table}[t]
\centering%
\caption{\small Computational complexity of model-order selection
algorithms.} {
\begin{tabular}{lcc}
\hline
{\small Algorithm} & {\small Additions} & {\small Multiplications} \\
\emph{\small \bf Auto-Rank}  & {\small $ 2(D_{max}- D_{min}) +1$}  & {\small $ -$}  \\
\emph{\small \bf (Extended Filters)}  & {\small $ 2(I_{max}-
I_{min}) +1$}  & {\small $ $}

\\\\
\emph{\small \bf Projection with } & {\small
$2(2M-1)\times $} & {\small $((M)^2+M+1)\times $} \\
\emph{\small \bf Stopping Rule }\cite{goldstein} & {\small
$(D_{max}- D_{min})+1$} & {\small $(D_{max}-D_{min}+1)$}
\\\\
\emph{\small \bf CV }\cite{avf5} & {\small $(2M-1)\times $} & {\small $(D_{max}- D_{min}+1)\times$} \\
\emph{\small \bf  } & {\small $(2(D_{max}- D_{min}) +1)$} &
{\small $M + 1$}
\\\\
\emph{\small \bf  } & {\small $2(D_{max}- D_{min}) +1$} & {\small $+ 7D_{max}^2+9D_{max}$} \\
\emph{\small \bf  Multiple Filters \cite{jidf}} & {\small $f_a(D_{max},I_{max}) + \ldots $} & {\small $f_m(D_{max},I_{max}) + \ldots $} \\
\emph{\small \bf  (JIDF or BARC) } & {\small $ + f_a(D_{min},I_{min})$} & {\small $+ f_m(D_{min},I_{min})$} \\
\emph{\small \bf   } & {\small $ 2(D_{max}- D_{min})
+1 $} & {\small $2(I_{max}- I_{min}) +1$} \\
%\emph{\small \bf  } & {\small $2(D_{max}- D_{min}) +1$} & {\small $(D_{max}-D_{min}+1)\times$} \\
%\emph{\small \bf  Multiple Filters \cite{jidf}} & {\small $(D_{max}-D_{min}+1)\times$} & {\small $(7(JM)^2 + 2JM$} \\
%\emph{\small \bf  (other schemes)} & {\small $ (3(JM)^2 - 2JM + 3 $} & {\small $+ 7D_{max}^2+9D_{max})$} \\
%\emph{\small \bf  } & {\small $ +6D_{max}^2 - 8D_{max} + 3)$} &
%{\small }\\
 \hline
\end{tabular}
}
\end{table}

The computational complexity of the proposed model-order selection
algorithm (Auto-Rank) and the existing rank selection algorithms
is shown in Table III. We can notice that the proposed model-order
selection algorithm with extended filters is significantly less
complex than the existing methods based on projection with
stopping rule \cite{goldstein} and the CV approach \cite{avf5}.
Specifically, the proposed rank selection algorithm with extended
filters only requires $2(D_{max}-D_{min}) + 2(I_{max}-I_{min})$
additions, as depicted in the first row of Table III, in addition
to the operations required by the proposed algorithms, whose
complexity is shown in the last rows of Tables I and II. For the
operation of the MSWF and the AVF algorithms with model-order
selection algorithms, a designer must add the complexities in
Tables I and II to the complexity of the model-order selection
algorithm of interest, as shown in Table III. The model-order
selection algorithm with multiple filters has a number of
arithmetic operations that is substantially higher than the other
compared methods and requires the computation of $(D_{\rm max} -
D_{\rm min}+1)+(I_{\rm max} - I_{\rm min}+1)$ pairs of filters
with costs $f_a(D,I)$ and $f_m(D,I)$ for additions and
multiplications, respectively, for each pair of filters with $D$
and $I$. Specifically, these costs are shown as a function of $D$
and $I$ at the bottom of Table III and we have for the SG version
$f_a(D,I)=4D+ BD +4I + (I-1)M-2$ additions and $f_m(D,I)=4D +
B(D+1)+ 5I+IM + 4$ multiplications (see the last rows of Table I),
whereas for the RLS version we have $f_a(D,I)=6D^2 +(B+1) D +6I^2
+I+ (I-1)M +8$ additions and $f_m(D,I)=7D^2 + (B+8)D + 7I^2 +7I+
IM +3$ multiplications (see the last rows of Table II). It .
Despite the cost, its performance is comparable with the proposed
model-order selection algorithm with extended filters. }

\section{Analysis of the Proposed Algorithms}

In this section, we develop a stability analysis of the proposed
method and SG algorithms and study the convergence issues of the
optimization problem. Specifically, we study the existence of
multiple solutions and discuss strategies for dealing with it. We
consider particular instances of the proposed algorithms for which
a global minimum may be encountered by the proposed SG and RLS
algorithms. We also examine cases for which there is no guarantee
that the algorithms will converge to the global minimum and may
end up in local minima. It should be mentioned, however, that the
proposed SG and RLS algorithms were extensively tested for a
number of applications and numerous scenarios. It was verified in
these experiments that the algorithms always converge to
approximately the same filter values irrespective of the
initialization. This suggests that the problem may have multiple
global minima or that every point of minimum is a point of global
minimum or that the switching of branches allows the algorithms to
find the global minimum.  Specifically, we are interested in
examining three cases of adaptation and parameter estimation,
namely:

\begin{itemize}
\item{Case i) - ${\boldsymbol S}_D[i]$ is fixed, i.e. the
interpolator ${\boldsymbol v}[i]$ and the decimation matrix
${\boldsymbol D}[i]$ are fixed.}

\item{Case ii) - ${\boldsymbol S}_D[i]$ is time-variant with
${\boldsymbol D}[i]$ being fixed and ${\boldsymbol v}[i]$ being
time-variant.}

\item{Case iii) - ${\boldsymbol S}_D[i]$ is time-variant, where
${\boldsymbol D}[i]$ and ${\boldsymbol v}[i]$ are both
time-variant.}

{  \item{Case iv) - ${\boldsymbol S}_D[i]$ is time-variant, where
${\boldsymbol D}[i]$ is time-variant and ${\boldsymbol v}[i]$ is
time-invariant.}}
\end{itemize}

For the sake of analysis and the convexity issues of the problem,
we have opted for studying the method for the four cases
previously outlined. This allows us to gain further insight and
draw conclusions on the properties of the different configurations
of the method. A key feature of the proposed method which makes
its convergence study extremely {  difficult} is the combined use
of discrete and continuous optimization techniques. Even though
the necessary conditions for the optimization algorithms are met
\cite{luen,kelley} and the cost functions used for deriving the SG
and RLS algorithms are continuously differentiable, the discrete
nature of the decimation and the patterns used make its
theoretical analysis highly challenging. This proof is beyond the
scope of this paper and remains a very interesting open problem.

\subsection{Stability Analysis}

In this part, we examine the stability of the proposed SG
algorithms. In order to establish these conditions, we define the
error matrices at time $i$ as
\begin{equation}
\begin{split}
{\boldsymbol E}_{{\boldsymbol S}_D}[i] & = {\boldsymbol S}_D[i] -
{\boldsymbol S}_{D,{\rm opt}} ~~ {\rm and} \\
{\boldsymbol e}_{\bar{\boldsymbol w}}[i] & = \bar{\boldsymbol
w}[i] - \bar{\boldsymbol w}_{\rm opt}, \label{errvecs}
\end{split}
\end{equation}
where $\bar{\boldsymbol w}_{\rm opt}$ and ${\boldsymbol S}_{D,{\rm
opt}}$ are the optimal parameter estimators. Since we are dealing
with a joint optimization procedure, both filters have to be
considered jointly. At this point, we need to introduce a
mathematical manipulation that allows the expression of
${\boldsymbol S}_D[i+1] = {\boldsymbol V}[i+1] {\boldsymbol
D}^H[i+1]$ as a function of the recursion in (\ref{sgrec}). We can
rewrite ${\boldsymbol S}_D[i+1]$ as
\begin{equation}
\begin{split}
{\boldsymbol S}_D[i+1] & = {\boldsymbol V}[i+1] {\boldsymbol
D}^H[i+1] = \sum_{l=1}^{M} {\boldsymbol B}_l {\boldsymbol v}[i+1]
{\boldsymbol D}^H[i+1] \\
& = \sum_{l=1}^{M} {\boldsymbol B}_l {\boldsymbol v}[i]
{\boldsymbol D}^H[i] - \mu_v e[i] z^*[i] \sum_{l=1}^{M}
{\boldsymbol B}_l ( {\boldsymbol I} - ({\boldsymbol
p}_{\bar{\boldsymbol w}}^H[i]{\boldsymbol p}_{\bar{\boldsymbol
w}}[i])^{-1} {\boldsymbol p}_{\bar{\boldsymbol w}}[i]{\boldsymbol
p}_{\bar{\boldsymbol w}}^H[i]) {\boldsymbol u}[i] {\boldsymbol
D}^H[i] \\
& = {\boldsymbol S}_D[i] - \mu_v e[i] z^*[i] {\boldsymbol B}_w[i]
{\boldsymbol u}[i] {\boldsymbol D}^H[i], \label{mrecs}
\end{split}
\end{equation}
where the $M \times I$ matrix ${\boldsymbol B}_w[i] =
\sum_{l=1}^{M} {\boldsymbol B}_l \bigg( {\boldsymbol I} -
({\boldsymbol p}_{\bar{\boldsymbol w}}^H[i]{\boldsymbol
p}_{\bar{\boldsymbol w}}[i])^{-1} {\boldsymbol
p}_{\bar{\boldsymbol w}}[i]{\boldsymbol p}_{\bar{\boldsymbol
w}}^H[i]\bigg)$, and the $M \times I$ matrix ${\boldsymbol B}_l$
has an $I$-dimensional identity matrix starting at the $l$-th row,
is shifted down by one position for each $l$ and the other
elements are zeros.

By substituting the expressions of ${\boldsymbol E}_{{\boldsymbol
S}_D}[i]$ and ${\boldsymbol e}_{\bar{\boldsymbol w}}[i]$ in
(\ref{mrecs}) and (\ref{wrecsgj}), respectively, and rearranging
the terms we obtain {
\begin{equation}
\begin{split}
{\boldsymbol E}_{{\boldsymbol S}_D}[i+1] & = \big\{ {\boldsymbol
I} - \mu_v e[i] {\boldsymbol B}_w[i] {\boldsymbol u}[i]
{\boldsymbol D}^H[i] {\boldsymbol r}^H[i] \big\}  {\boldsymbol
E}_{{\boldsymbol S}_D}[i]
\\ & \quad - \mu_v e[i] {\boldsymbol B}_w[i] {\boldsymbol u}[i]
{\boldsymbol D}^H[i] {\boldsymbol r}^H[i] {\boldsymbol S}_D[i]
{\boldsymbol e}_{\bar{\boldsymbol w}}[i] \\
& \quad + \mu_v e[i] {\boldsymbol B}_w[i] {\boldsymbol u}[i]
{\boldsymbol D}^H[i] {\boldsymbol r}^H[i] ({\boldsymbol S}_D[i]
({\boldsymbol I} - \bar{\boldsymbol w}_{\rm opt}) - {\boldsymbol
S}_{D, {\rm opt}} ), \label{esd}
\end{split}
\end{equation}
{  \begin{equation}
\begin{split}
{\boldsymbol e}_{\bar{\boldsymbol w}}[i+1] & = \big\{ {\boldsymbol
I} - \mu_w e[i] {\boldsymbol \Pi}[i] {\boldsymbol S}_D^H[i]
{\boldsymbol r}[i] {\boldsymbol r}^H[i] {\boldsymbol S}_D[i]
\big\} {\boldsymbol e}_{\bar{\boldsymbol w}}[i] \\
& \quad  - \mu_w e[i] {\boldsymbol \Pi}[i] {\boldsymbol S}_D^H[i]
{\boldsymbol r}[i] {\boldsymbol r}^H[i] {\boldsymbol
E}_{{\boldsymbol S}_D}[i] \\ & \quad - \mu_w e[i] {\boldsymbol
\Pi}[i] {\boldsymbol S}_D^H[i] {\boldsymbol r}[i] {\boldsymbol
r}^H[i] ({\boldsymbol E}_{{\boldsymbol S}_D}[i] (\bar{\boldsymbol
w}_{\rm opt} - {\boldsymbol I}) + {\boldsymbol S}_{D,{\rm
opt}}\bar{\boldsymbol w}_{\rm opt} ), \label{ew}
\end{split}
\end{equation}}}
where ${\boldsymbol \Pi}[i] = {\boldsymbol I} - (\bar{\boldsymbol
p}^H[i] \bar{\boldsymbol p}[i] )^{-1} \bar{\boldsymbol p}[i]
\bar{\boldsymbol p}^H[i]$. Taking expectations and considering the
two error matrices together, we obtain
\begin{equation}
\begin{split}
\left[\begin{array}{c}
  E\big[{\boldsymbol E}_{{\boldsymbol S}_D}[i+1]\big] \\
  E\big[{\boldsymbol e}_{\bar{\boldsymbol w}}[i+1]\big]  ~|~ {\boldsymbol 0}_{D \times (M-1)}
\end{array}\right] & = {\boldsymbol A}
\left[\begin{array}{c}
  E\big[{\boldsymbol E}_{{\boldsymbol S}_D}[i]\big] \\
  E\big[{\boldsymbol e}_{\bar{\boldsymbol w}}[i]\big] ~|~ {\boldsymbol 0}_{D \times (M-1)}
\end{array}\right] + {\boldsymbol C}, \label{stab}
 \end{split}
\end{equation}
where \begin{equation}{\boldsymbol A} =  \left[ \hspace{-0.5em}
\begin{array}{c c}
  {\small \big\{ {\boldsymbol
I} - \mu_v e[i] {\boldsymbol B}_w[i] {\boldsymbol u}[i]
{\boldsymbol D}^H[i] {\boldsymbol r}^H[i] \big\}} & {\small \mu_v
e[i] {\boldsymbol B}_w[i] {\boldsymbol u}[i]
{\boldsymbol D}^H[i] {\boldsymbol r}^H[i] {\boldsymbol S}_D[i]} \\
  {\small \big\{ {\boldsymbol I}
- \mu_w e[i] {\boldsymbol \Pi}[i] {\boldsymbol S}_D^H[i]
{\boldsymbol r}[i] {\boldsymbol r}^H[i] {\boldsymbol S}_D[i]
\big\} }  & {\small - \mu_w e[i] {\boldsymbol \Pi}[i] {\boldsymbol
S}_D^H[i] {\boldsymbol r}[i] {\boldsymbol r}^H[i]}
\end{array} \hspace{-0.5em} \right],\nonumber\end{equation}
\begin{equation}{\boldsymbol C} = \left[\hspace{-0.5em} \begin{array}{c}
{\small  + \mu_v e[i] {\boldsymbol B}_w[i] {\boldsymbol u}[i]
{\boldsymbol D}^H[i] {\boldsymbol r}^H[i] ({\boldsymbol S}_D[i]
({\boldsymbol I} - \bar{\boldsymbol w}_{\rm opt}) - {\boldsymbol
S}_{D, {\rm opt}} )} \\
{\small  - \mu_w e[i] {\boldsymbol \Pi}[i] {\boldsymbol S}_D^H[i]
{\boldsymbol r}[i] {\boldsymbol r}^H[i] ({\boldsymbol
e}_{{\boldsymbol S}_D}[i] (\bar{\boldsymbol w}_{\rm opt} -
{\boldsymbol I}) + {\boldsymbol S}_{D,{\rm opt}}\bar{\boldsymbol
w}_{\rm opt} ) ~ {\boldsymbol 0}_{D \times (M-1)}}
\end{array}\hspace{-0.5em} \right].\nonumber
\end{equation}
{  The previous equations imply that the stability of the
algorithms depends on the spectral radius of ${\boldsymbol A}$.
The parameters of $\bar{\boldsymbol w}[i]$ and ${\boldsymbol
S}_D[i]$ will remain bounded and will converge asymptotically to
the optimal values if the step sizes are chosen such the
eigenvalues of ${\boldsymbol A}^H{\boldsymbol A}$ are less than
one.} Unlike the stability analysis of most adaptive algorithms
\cite{haykin}, in the proposed approach the terms are more
involved and depend on each other as evidenced by the equations
for ${\boldsymbol A}$ and ${\boldsymbol C}$. Let us now examine
the three cases outlined at the beginning of this section.

For case i), the transformation ${\boldsymbol S}_D$ is fixed and
we can consider only the recursion for the error vector
${\boldsymbol e}_{\bar{\boldsymbol w}}[i]$, which yields
\begin{equation}
\begin{split}
{\boldsymbol e}_{\bar{\boldsymbol w}}[i+1] & = ( {\boldsymbol I} -
\mu_w e[i] {\boldsymbol \Pi}[i] {\boldsymbol S}_D {\boldsymbol
r}[i] {\boldsymbol r}^H[i] {\boldsymbol S}_D ) {\boldsymbol
e}_{\bar{\boldsymbol w}}[i] \\ & \quad - \mu_w e[i] {\boldsymbol
\Pi}[i] {\boldsymbol S}_D^H {\boldsymbol r}[i] {\boldsymbol
r}^H[i] {\boldsymbol S}_D \bar{\boldsymbol w}_{\rm opt}.
\end{split}
\end{equation}
Taking expectations on both sides, using the fact that
$E\big[{\boldsymbol e}_{\bar{\boldsymbol w}}[i] \big] =
{\boldsymbol 0}$ and computing ${\boldsymbol R}_{\bar{\boldsymbol
w}} = E\big[{\boldsymbol e}_{\bar{\boldsymbol w}}[i]{\boldsymbol
e}_{\bar{\boldsymbol w}}^H[i]\big]$ we get
\begin{equation}
\begin{split}
{\boldsymbol R}_{\bar{\boldsymbol w}} & = ( {\boldsymbol I} -
\mu_w E[e[i] {\boldsymbol \Pi}[i]] {\boldsymbol S}_D^H
{\boldsymbol R} {\boldsymbol S}_D^H ){\boldsymbol
R}_{\bar{\boldsymbol w}} ( {\boldsymbol I} - \mu_w E[e[i]
{\boldsymbol \Pi}[i]] {\boldsymbol S}_D^H {\boldsymbol R}
{\boldsymbol r}^H[i] {\boldsymbol S}_D ) \\ & \quad \mu_w^2
E[|e[i]|^2{\boldsymbol \Pi}[i]] {\boldsymbol S}_D {\boldsymbol R}
{\boldsymbol S}_D  \bar{\boldsymbol w}_{\rm opt}^H {\boldsymbol
S}_D^H {\boldsymbol R} {\boldsymbol S}_D E[{\boldsymbol
\Pi}^H[i]],
\end{split}
\end{equation}
where ${\boldsymbol R} = E [ {\boldsymbol r}[i] {\boldsymbol
r}^H[i]]$ is the $M \times M$ covariance matrix of the input
${\boldsymbol r}[i]$. Using well-known results from the theory in
\cite{haykin}, we have the following stability condition
\begin{equation}
0 < \mu_w < \frac{2}{tr\Big[E\big[e[i]{\boldsymbol
\Pi}[i]\big]\Big] {\boldsymbol S}_D^H {\boldsymbol R} {\boldsymbol
S}_D  }
\end{equation}
For case ii) we assume that ${\boldsymbol D}[i]$ is fixed and
${\boldsymbol v}[i]$ and $\bar{\boldsymbol w}[i]$ are
time-variant, which means the trajectories of ${\boldsymbol
S}_D[i]$ and $\bar{\boldsymbol w}[i]$ must be considered jointly.
Therefore, the equation in (\ref{stab}) should be used in the
analysis. For stability, the step sizes should be adjusted such
that the eigenvalues of ${\boldsymbol A}^H{\boldsymbol A}$ are
less than one. Despite this condition of stability the algorithms
may converge to local minima. In what follows, we will study this.

{  For cases iii) and iv), we consider that ${\boldsymbol D}[i]$,
${\boldsymbol v}[i]$ and $\bar{\boldsymbol w}[i]$ are time-variant
and ${\boldsymbol D}[i]$ and $\bar{\boldsymbol w}[i]$ are
time-variant, respectively.} The condition of stability is
different from the previous cases since ${\boldsymbol D}[i]$ is a
discretely optimized parameter and ${\boldsymbol v}[i]$ and
$\bar{\boldsymbol w}[i]$ are parameter vectors that are
continuously optimized. The equation in (\ref{stab}) still holds
but the discrete nature of ${\boldsymbol D}[i]$ makes a precise
stability analysis impractical since ${\boldsymbol D}[i]$ is
switched every time instant. In addition, the problem becomes very
difficult to treat since local minima may arise due to the joint
adaptation of ${\boldsymbol D}[i]$, ${\boldsymbol v}[i]$ and
$\bar{\boldsymbol w}[i]$ (case iii)) and the joint adaptation of
${\boldsymbol D}[i]$ and $\bar{\boldsymbol w}[i]$ (case iv)).

\subsection{Analysis of the Optimization Problem}

Let us now consider an analysis of the joint optimization method
from the point of view of the cost function and the constraints.
Our strategy is to examine the four cases previously outlined and
draw conclusions on what happens to the nature of the optimization
problem. Let us drop the time index $[i]$ for simplicity and
define the cost function
\begin{equation}
\begin{split}
J_{\rm CM}({\boldsymbol v},{\boldsymbol D} ,\bar{\boldsymbol w}) &
= E\Big[\big(|\bar{\boldsymbol w}^{H}{\boldsymbol S}_D
{\boldsymbol r} |^{2} - 1 \big)^2\Big]\\ & =
E\Big[\big(|\bar{\boldsymbol w}^{H}{\boldsymbol D}
\boldsymbol{\Re}_o{\boldsymbol v}^{*} |^{2} - 1 \big)^2\Big]\\
& = E\Big[\big(|{\boldsymbol t}^{H}{\boldsymbol U} {\boldsymbol t}
|^{2} - 1 \big)^2\Big] \\ & = E\Big[ |z|^4 - 2 |z|^2 +1],
\label{barccost2}
\end{split}
\end{equation}
where the $(D+I) \times 1$ parameter vector ${\boldsymbol t} = [
\bar{\boldsymbol w}^T ~ {\boldsymbol v}^T]^T$ considers together
the reduced-rank estimator and the interpolator and the $(D+I)
\times (D+I)$ matrix ${\boldsymbol U} = \left[ \begin{array}{cc}
{\boldsymbol 0} & {\boldsymbol 0} \\ {({\boldsymbol D}
{\boldsymbol \Re}_o)^T } & {\boldsymbol 0} \end{array} \right]$
contains the samples of the received vector and the decimation
matrix.

The received vector in (\ref{recsignal}) can be rewritten as
${\boldsymbol r} = {\boldsymbol x} + {\boldsymbol \eta} +
{\boldsymbol n}$, where ${\boldsymbol x}= \sum_{k=1}^{K}A_{k}
{b}_{k}{\boldsymbol p}_{k}$ and ${\boldsymbol p}_{k}={\boldsymbol
C}_{k}{\boldsymbol h}_k$. Since the symbols $b_{k}$,
$k=1,\ldots,K$ are i.i.d. complex random variables with mean zero
and unit variance, $b_{k}$ and ${\boldsymbol n}$ are statistically
independent, and we have ${\boldsymbol R}={\boldsymbol R}_x + {\bf
R}_{\eta} +  \sigma^{2}{\bf I}$, where ${\boldsymbol
R}_x=E[{\boldsymbol x}{\boldsymbol x}^{H}]$ and ${\boldsymbol
R}_{\eta}=E[\boldsymbol{\eta}\boldsymbol{\eta}^{H}]$.

Let us consider a desired user and its corresponding
transformation matrix ${\boldsymbol S}_{D}$ and reduced-rank
estimator $\bar{\boldsymbol w}$. We can express the interference
free desired signal as
\begin{equation}
q_{k}=A_{k}{\boldsymbol p}^{H} {\boldsymbol S}_{D}\bar{\boldsymbol
w}
\end{equation}
and the composite signal as
\begin{equation}
{\boldsymbol q}={\boldsymbol A} \big[{\boldsymbol p}_1, ~
{\boldsymbol p}_2,~ \ldots,~  {\boldsymbol p}_K \big]^{H}
{\boldsymbol S}_{D}\bar{\boldsymbol w} = {\boldsymbol A}
{\boldsymbol P}^H {\boldsymbol S}_{D}\bar{\boldsymbol w},
\end{equation}
where ${\boldsymbol A}={\rm diag}(A_{1} \ldots A_{K})$ is a $K
\times K$ diagonal matrix with the amplitudes, ${\boldsymbol
P}=[{\boldsymbol p}_{1} \ldots {\boldsymbol p}_{K}]$ is a $M\times
K$ matrix with the effective signatures.

Now let us make use of the constraint $\bar{\boldsymbol w}^{H}
{\boldsymbol S}_{D} {\boldsymbol p}_k= \bar{\boldsymbol
w}^{H}{\boldsymbol S}_{D} {\boldsymbol C}_k {\boldsymbol h}_k =
\nu $ and the relation between ${\boldsymbol S}_{D}$,
$\bar{\boldsymbol w}$, the channel and the signature ${\boldsymbol
C}_k^H {\boldsymbol S}_{D}\bar{\boldsymbol w} = \nu
\hat{\boldsymbol h}_k$ \cite{xutsa,xu&liu,mswfccm}. We then have
for the desired user the equivalent expressions
\begin{equation}
\begin{split}
q_{k} & =A_{k}{\boldsymbol p}_{k}^{H}{\boldsymbol
S}_{D}\bar{\boldsymbol w} = A_{k}{\boldsymbol h}_1^H{\boldsymbol
C}_{k}^{H}\bar{\boldsymbol w}= \nu A_{k}{\boldsymbol
h}_k^{H}\hat{\boldsymbol h}_k \\
&  = A_{k}{\boldsymbol p}_{k}^{H}{\boldsymbol V} {\boldsymbol D}^H
\bar{\boldsymbol w} = A_k {\boldsymbol v}^H  {\boldsymbol \Re}_p^H
{\boldsymbol D}^H \bar{\boldsymbol w} = A_k {\boldsymbol t}^H
{\boldsymbol U}_p^H {\boldsymbol t}, \label{equiv}
\end{split}
\end{equation}
where the $(D+I) \times (D+I)$ matrix ${\boldsymbol U}_p = \left[
\begin{array}{cc} {\boldsymbol 0} & {\boldsymbol 0} \\
{({\boldsymbol D} {\boldsymbol \Re}_p)^T } & {\boldsymbol 0}
\end{array} \right]$ and the $M \times I$ Hankel matrix ${\boldsymbol \Re}_p
$ contains shifted versions of the effective signature
${\boldsymbol p}_k$ of the desired user.

At this point, we can exploit the previous expressions and
substitute them into the cost function in (\ref{barccost2}).
Assuming for simplicity the absence of noise and ISI, the cost
function of the desired signal can be expressed as
\begin{equation}
\begin{split}
J_{CM}({\boldsymbol q}) & = E[({\boldsymbol q}^{H}{\boldsymbol
b}{\boldsymbol b}^{H}{\boldsymbol q})^{2}] - 2E[({\boldsymbol
q}^{H}{\boldsymbol b}{\boldsymbol b}^{H}{\boldsymbol q})] +1 \\ &
= 8 ( F + \sum_{l=2}^{K} q_{l}q_{l}^{*})^{2} - 4F^{2} -
4\sum_{l=2}^{K} (q_{l}q_{l}^{*})^{2} - 4F -4 \sum_{l=2}^{K}
(q_{l}q_{l}^{*}) + 1, \label{costmod}
\end{split}
\end{equation}
where $F = q_{k}q_{k}^{*}= A_k^2 | {\boldsymbol t}^H {\boldsymbol
U}_p^H {\boldsymbol t}|^2 =\nu^{2} A_{k}^{2}|\hat{\boldsymbol
h}_k^{H}{\boldsymbol h}_k|^{2}$ and ${\boldsymbol b} =[b_{1}
\ldots b_{K}]^{T}$ is a $K \times 1$ vector with the transmitted
symbols.

In order to study the properties of the optimization of
(\ref{costmod}), we proceed as follows. We take advantage of the
constraint $\bar{\boldsymbol w}^{H} {\boldsymbol S}_{D}
{\boldsymbol p}_k = \nu$ and rewrite (\ref{costmod}) as
\begin{equation}
\tilde{J}_{CM}(\bar{\boldsymbol q})= 8 ( F + \bar{\boldsymbol
q}^{H}\bar{\boldsymbol q})^{2} - 4(F^{2} +
\sum_{l=2}^{K}(q_{l}q_{l}^{*})^{2}) - 4(F + \bar{\boldsymbol
q}^{H}\bar{\boldsymbol q}) + 1,
\end{equation}
where $\bar{\boldsymbol q}=[q_{2},\ldots, q_{K}]^{T}={\boldsymbol
T}{\boldsymbol S}_D\bar{\boldsymbol w}$, ${\boldsymbol T}=
{\boldsymbol A}'^{H}{\boldsymbol P}'^{H}$, ${\boldsymbol
P}'=[{\boldsymbol p}_{2} \ldots {\boldsymbol p}_{K}]$ and
${\boldsymbol A}'={\rm diag}(A_{2} \ldots A_{K})$.

The previous development allows us to examine the four cases
outlined at the beginning of the section via the computation of
the Hessian matrix (${\boldsymbol \Theta}$) using ${\boldsymbol
\Theta} = \frac{\partial }{\partial \bar{\boldsymbol q}^{H}}
\frac{\partial (\tilde{J}_{CM}(\bar{\boldsymbol q}))}{\partial
\bar{\boldsymbol q}}$. Specifically, ${\boldsymbol \Theta}$ is
positive definite if ${\boldsymbol m}^{H}{\boldsymbol
\Theta}{\boldsymbol m}> 0$ for all nonzero ${\boldsymbol m} \in
\boldsymbol{C}^{K-1\times K-1}$ \cite{golub}. The computation of
${\boldsymbol \Theta}$ is given by
\begin{equation}
{\boldsymbol \Theta} = 16\Big[  (F -1/4){\boldsymbol I} +
\bar{\boldsymbol q}^{H}\bar{\boldsymbol q}{\boldsymbol I} +
\bar{\boldsymbol q}\bar{\boldsymbol q}^{H} - {\rm
diag}(|q_{2}|^{2} \ldots |q_{K}|^{2}) \Big], \label{hess}
\end{equation}
where the first term depends on $F$ and the selection of some key
parameters, the second term is positive definite, and the third
and fourth terms of (\ref{hess}) are positive semi-definite
matrices. We will now consider the four cases of interest for our
analysis.

For case i), we assume ${\boldsymbol S}_D$ fixed and $F$ yields
the condition
\begin{equation}
\nu^{2} A_{k}^{2}|\hat{\boldsymbol h}_k^{H}{\boldsymbol h}_k|^{2}
\geq 1/4,
\end{equation}
that ensures the convexity of the optimization problem in the
noiseless case. Since $\bar{\boldsymbol q}={\boldsymbol T}
{\boldsymbol S}_D\bar{\boldsymbol w}$ is a linear mapping of
${\boldsymbol S}_D$ and $\bar{\boldsymbol w}$, then
$\tilde{J}_{CM}(\bar{\boldsymbol q})$ is a convex function of
$\bar{\boldsymbol q}$ and implies that ${J}_{CM}( {\boldsymbol
S}_D, {\boldsymbol w}) = \tilde{J}_{CM}({\boldsymbol
T}{\boldsymbol S}_D\bar{\boldsymbol w})$ is a convex function of
${\boldsymbol S}_D\bar{\boldsymbol w}$.

For case ii), we suppose that ${\boldsymbol S}_D$ is time-variant
due to the interpolator ${\boldsymbol v}$ and we shall consider
${\boldsymbol v}$ and $\bar{\boldsymbol w}$ jointly via the
parameter vector ${\boldsymbol t}$. In this case, $F$ yields the
condition
\begin{equation}
A_k^2 | {\boldsymbol t}^H {\boldsymbol U}_p^H {\boldsymbol t}|^2
\geq 1/4,
\end{equation}
Although the optimization problem depends on the parameters
${\boldsymbol v}$ and $\bar{\boldsymbol w}$ which suggests a
nonconvex problem, there is the possibility of modifying the
problem with the condition above. As the extrema of the cost
function can be considered for small $\sigma^{2}$ a slight
perturbation of the noise-free case \cite{kwak}, the cost function
is also convex for small $\sigma^{2}$ provided the above
conditions hold.

For case iii), we assume that ${\boldsymbol D}$, ${\boldsymbol v}$
and $\bar{\boldsymbol w}$ are time-variant. The discrete nature of
${\boldsymbol D}$ and the switching between branches are clearly
associated with a nonconvex problem for which there is no easy or
known strategy to enforce convexity. Interestingly, the switching
does not affect the final values of the parameter vectors
${\boldsymbol v}$ and $\bar{\boldsymbol w}$ which converge to the
same steady state values regardless of the initialization,
provided ${\boldsymbol v}$ and ${\boldsymbol S}_D$ are not
all-zero quantities.

{  For case iv), we consider that ${\boldsymbol v}$ is
time-invariant, and ${\boldsymbol D}$ and $\bar{\boldsymbol w}$
are time-variant. The discrete nature of ${\boldsymbol D}$ and the
switching between branches are again associated with a nonconvex
problem for which there is no simple strategy to enforce
convexity. An analysis of this problem for cases iii) and iv)
remains an interesting open problem.}

\section{Simulations}

In this section we evaluate the bit error rate (BER) performance
of the proposed BARC scheme and algorithms in a DS-CDMA
interference suppression application. We consider the system model
detailed in Section II and model the channel as a finite impulse
response (FIR) filter represented as the $L_p \times 1$ channel
vector ${\boldsymbol h}_k[i] = [{h}_{k,0}[i] ~\ldots ~
{h}_{k,L_{p}-1}[i]]^T$ \cite{rappa} The system employs random
sequences of length $N=32$ and $N=64$. All the multipath channels
are time-varying and are generated according to Clarke's model
\cite{rappa}, which is parameterized by the normalized Doppler
frequency $f_D T$, where $f_D$ is the Doppler frequency and $T$ is
the inverse of the symbol rate. We assume $L_{p}=9$ as an upper
bound, which means ${\boldsymbol r}[i]$ has $M= N+L_p-1= 40$ when
$N=32$ and $M=72$ taps when $N=64$, respectively. In this case,
the ISI corresponds to $3$ symbols namely, the current, previous
and successive symbols. In all simulations, we assume $L_p=9$ as
an upper bound, $3$-path channels with relative powers given by
$0$, $-3$ and $-6$ dB, where in each run the spacing between paths
is obtained from a discrete uniform random variable between $1$
and $2$ chips and we average the curves over $200$ runs. The
system has a power distribution among the users for each run that
follows a log-normal distribution with standard deviation equal to
$1.5$ dB. {  The blind algorithms employ the CCM criterion,
adaptive linear receivers that assume perfect synchronization and
know the spreading code of the user of interest. The number of
users $K$ does not affect the complexity of a receiver designed
for a particular user.} We measure the BER of the desired user and
compare the BARC scheme with the full-rank
\cite{xu&liu},\cite{delamareccm}, reduced-rank schemes with the
MSWF method \cite{mswfccm}, the AVF scheme with training
\cite{avf5}, the JIO technique \cite{jio} and the SVD-based
approach that selects the $D$ largest eigenvectors
\cite{wang&poor} to compute the transformation matrix
${\boldsymbol S}_D[i]$ and the MMSE, which assumes the knowledge
of the channels and the noise variance. All algorithms have their
parameters optimized with respect to the BER for each scenario and
the blind algorithms employ the blind channel estimator of
\cite{douko} to compute the effective signature ${\boldsymbol
p}[i]$. The phase ambiguity derived from the blind channel
estimation method in \cite{douko} is eliminated in our simulations
by using the phase of { $\hat{\bf h}_k[0]$} as a reference to
remove the ambiguity.

\subsection{Model-Order Adjustment}

In most estimation algorithms, it is necessary to adjust
parameters such as order, step size and forgetting factor. In the
proposed BARC scheme, a key issue is the setting of the number of
elements or the rank of the estimators ${\boldsymbol v}[i]$ and
$\bar{\boldsymbol w}[i]$ used. We have conducted experiments in
order to obtain the most adequate rank for the interpolator
${\boldsymbol v}[i]$, with values ranging from $3$ to $8$ and for
the reduced-rank filter $\bar{\boldsymbol w}[i]$ with values
ranging from $1$ to $16$. Notice that values beyond that range are
unnecessary since it does not lead to performance improvements.

\begin{figure}[!htb]
\begin{center}
\def\epsfsize#1#2{0.9\columnwidth} \epsfbox{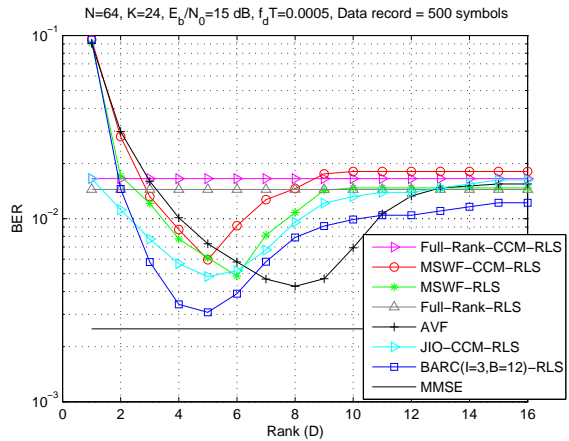}
\vspace{-0.75em}\caption{BER performance against rank (D) for the
analyzed schemes using RLS algorithms.} \label{sxr1}
\end{center}
\end{figure}

\begin{figure}[!htb]
\begin{center}
\def\epsfsize#1#2{1\columnwidth} \epsfbox{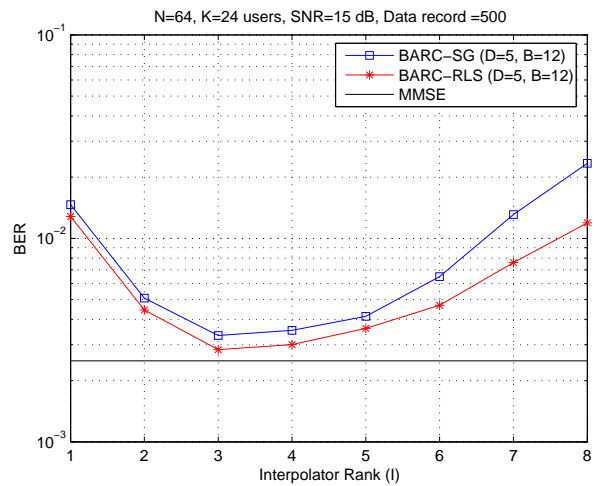}
\vspace{-0.75em}\caption{BER performance against interpolator rank
($I$) for the analyzed schemes using SG and RLS algorithms
$f_dT=0.0005$.} \label{sxr2}
\end{center}
\end{figure}

The results in Figs. \ref{sxr1} and \ref{sxr2} for a wide range of
scenarios indicate that the performance is good for a small range
of the number of taps in ${\boldsymbol v}[i]$ and
$\bar{\boldsymbol w}[i]$. While the BARC scheme is not able to
construct an appropriate subspace projection with only a few
coefficients in ${\boldsymbol v}(i)$ and $\bar{\boldsymbol w}(i)$,
there is no improvement in the tradeoff between model bias and
noise variance and the estimation task becomes slower when the
length of the estimator is greater than $6$. Thus, for this reason
and to keep a low complexity we adopt $I = 3$ and $D=5$ for the
next few experiments since these values yield the best
performance.

\subsection{Impact of Number of Branches and Decimation Schemes}

In this part, we evaluate the performance of the proposed BARC
scheme and algorithms for different decimation schemes, and the
impact of the number of branches on the performance.

\begin{figure}[!htb]
\begin{center}
\def\epsfsize#1#2{1\columnwidth}
\epsfbox{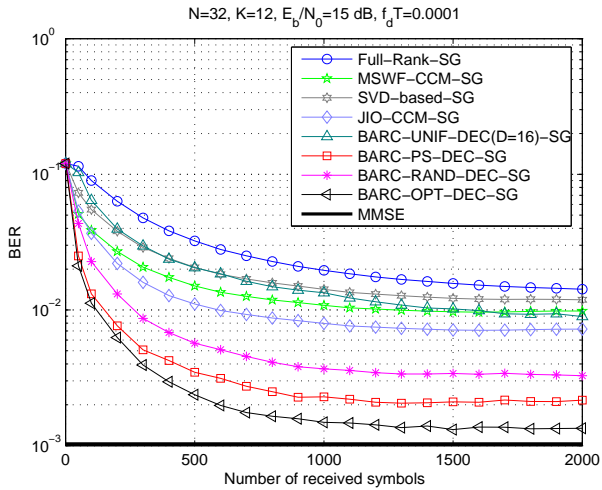} \vspace{-1em}
\vspace{-0.75em}\caption{\footnotesize BER performance versus number
of received symbols.} \label{dec}
\end{center}
\end{figure}

In order to assess the proposed decimation methods, we compute the
BER performance of the algorithms for the uniform (U-DEC), the
random (R-DEC), the prestored (PS-DEC) and the optimal (OPT-DEC)
schemes. The results, shown in Fig. \ref{dec}, indicate that the
BARC scheme with the optimal decimation (OPT-DEC) achieves the
best performance, followed by the proposed method with prestored
decimation (PS-DEC), the random decimation system (R-DEC), the
uniform decimation (U-DEC), the MSWF, the SVD and the full-rank
approach. Due to its exponential complexity, the optimal
decimation algorithm is not practical and the PS-DEC is the one
with the best trade-off between performance and complexity.

\begin{figure}[!htb]
\begin{center}
\def\epsfsize#1#2{1\columnwidth}
\epsfbox{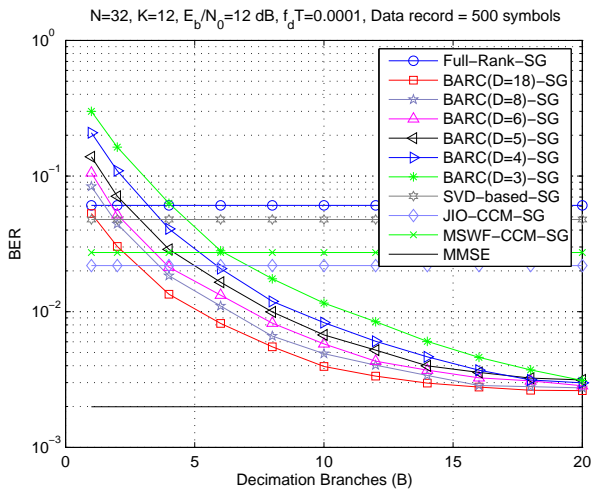} \vspace{-1em}\caption{\footnotesize BER
performance versus number of decimation branches.} \label{bb}
\end{center}
\end{figure}

In the next experiment, we evaluate the effect of the number of
decimation branches $B$ on the performance for various ranks $D$
with a data support of $1500$ symbols and the PS-DEC decimation
approach. The results, depicted in Fig. \ref{bb}, indicate that
the performance of the BARC scheme improves as $B$ is increased
and approaches the optimal MMSE estimator, which assumes that the
channels and the noise variance are known.

\subsection{Performance with Model-Order Selection}

\begin{figure}[!htb]
\begin{center}
\def\epsfsize#1#2{1\columnwidth}
\epsfbox{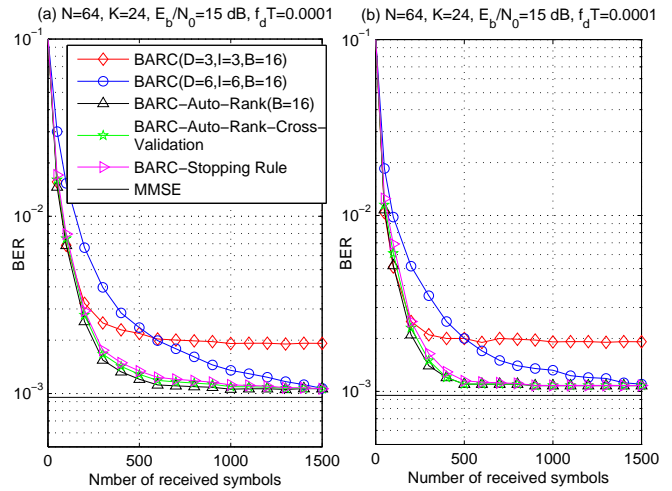} \vspace{-0.75em} \caption{\small BER performance
against number of symbols for different model-order selection
algorithms with (a) SG and (b) RLS recursions.} \label{rank}
\end{center}
\end{figure}

\begin{figure}[!htb]
\begin{center}
\def\epsfsize#1#2{1\columnwidth}
\epsfbox{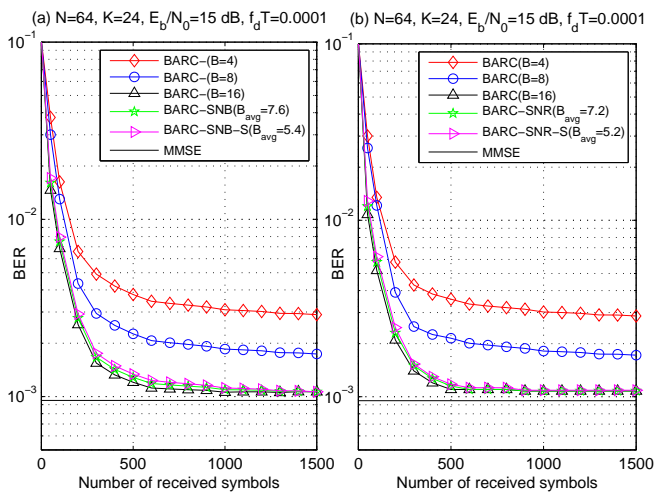} \caption{\small SINR performance against number
of symbols with (a) SG and (b) RLS recursions.} \label{branch}
\end{center}
\end{figure}

In the next experiments, shown in Figs. \ref{rank} and
\ref{branch}, we assess the performance of the BARC scheme with
the proposed model-order selection algorithm and mechanisms to
determine the minimum number of branches necessary to attain a
predefined performance as described in Section VI.

{  The evaluation of the model-order selection algorithms is shown
in Fig. \ref{rank}, where we consider the BARC scheme with SG and
RLS algorithms, $B=16$, $D_{\rm min}= 3$, $D_{\rm max}=6$, and
${I_{\rm min}}=2$ and ${I_{\rm max}}=6$. We compare a
configuration of the BARC scheme with $I=2$ and $D=3$, a second
configuration of the BARC with $I=6$ and $D=6$, the BARC with the
proposed model-order selection algorithm (Auto-Rank) with extended
filters, the BARC with the method based on the stopping rule of
\cite{goldstein} and the BARC with the CV-based algorithm of
\cite{avf5}. Notice that the BARC with the model-order selection
algorithm based on multiple filters obtains a comparable
performance to the Auto-Rank approach (the curves overlap and for
this reason we do not shot it), however, the former is
significantly more complex. The results indicate that the
Auto-Rank allows the BARC scheme to achieve fast convergence and
excellent steady state performance, which is close to the optimal
MMSE. The performance of the Auto-Rank is slightly better than the
stopping rule approach of \cite{goldstein} and the CV-based
technique of \cite{avf5}.   The proposed Auto-Rank algorithm is
less complex than the algorithms of \cite{goldstein} and
\cite{avf5} as it reduces the number of possible ranks to be used
by the estimators by constraining them in a preselected range and
does not require the computation of orthogonal projections as in
\cite{goldstein}.}

In the next experiment, we assess the proposed SNB and SNB-S
algorithms for automatically selecting the necessary number of
branches to attain a predefined performance.  The results are
shown in Fig. \ref{branch} for an identical scenario to Fig.
\ref{rank}. We consider the BARC scheme with SG and RLS algorithms
and the Auto-Rank algorithm for different values of $B$, and the
proposed SNR and SNR-S algorithms. The parameter $\rho$ was set
equal to $4\%$ greater than the MMSE and $B_{\rm{max}}=16$ for the
experiment. The results indicate that the proposed branch
adaptation techniques allow the BARC scheme to achieve a
performance comparable to the BARC scheme with $B=16$. In
particular, the proposed SNB algorithm achieves this performance
with $B_{\rm{avg}}=7.6$, whereas the proposed SNB-S technique
attains this performance with $B_{\rm{avg}}=5.4$ due to the use of
{\it a priori} knowledge of the frequency of branch usage. In the
following example, we consider the model-order selection and SNB-S
algorithms for the BARC with the same parameters used in the
previous experiment and the rank adaptation mechanisms proposed in
\cite{goldstein} for the MSWF and in \cite{avf5} for the AVF.

\subsection{Performance with Different Loads and SNR Values}

\begin{figure}[!htb]
\begin{center}
\def\epsfsize#1#2{1\columnwidth}
\epsfbox{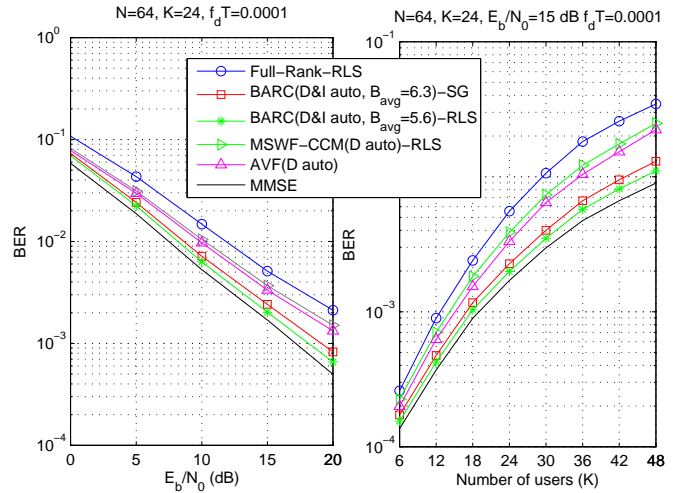} \caption{BER performance versus (a) $E_{b}/N_{0}$
(b) number of users for a data record of $500$ symbols.}
\label{bersu}
\end{center}
\end{figure}

In the last experiment, we assess the schemes and algorithms by
computing the BER performance against $E_{b}/N_{0}$ and the number
of users, as depicted in Fig. \ref{bersu}. The BER is evaluated
for data records of $1500$ QPSK symbols and a scenario where the
trained receivers employ pilot signals for estimating their
parameters with SG and RLS algorithms, whereas the blind
algorithms operate without any assistance. The maximum number of
branches for the BARC scheme is $B_{\rm max}=16$ and we employed
the proposed SNR-S algorithm.

The results show that the BARC scheme with both SG and RLS
algorithms achieves a BER performance very close to the optimal
MMSE, that assumes known channels, is followed by the AVF, the
MSWF-RLS and the full-rank. Specifically, the BARC scheme with the
SG algorithm
 can save up to $4$ dB in $E_b/N_0$ as compared to the AVF and
the MSWF-RLS for the same BER and can accommodate up to $6$ more
users as compared to the AVF and the MWF-RLS for the same BER.

\section{Conclusions}

This work proposes the BARC scheme and blind adaptive algorithms
for interference suppression in wireless communications systems.
The proposed BARC scheme employs a reduced-rank decomposition
based on the concept of joint interpolation, switched decimation
and reduced-rank estimation subject to a set of constraints. The
proposed set of constraints ensures that the multi-path components
of the channel are combined prior to dimensionality reduction. {
We have developed low-complexity SG and RLS reduced-rank
estimation and model-order selection algorithms along with
techniques for determining the required number of switching
branches to attain a predefined performance. We have applied the
proposed algorithms to interference suppression in DS-CDMA
systems.} The results of simulations indicate that the proposed
BARC scheme allows a substantially better convergence and tracking
performance than existing reduced-rank and full-rank schemes. This
is due to the dimensionality reduction carried out by the proposed
scheme that allows the use of adaptive algorithms with very small
estimators. The proposed algorithms can be applied to other
applications including MIMO systems, beamforming, broadband
channel equalization and navigation systems.

\begin{biography}{Rodrigo C. de Lamare}
(S'99 - M'04 - SM'10) received the Diploma in electronic
engineering from the Federal University of Rio de Janeiro (UFRJ)
in 1998 and the M.Sc. and PhD degrees, both in electrical
engineering, from the Pontifical Catholic University of Rio de
Janeiro (PUC-Rio) in 2001 and 2004, respectively. Since January
2006, he has been with the Communications Research Group,
Department of Electronics, University of York, where he is
currently a lecturer in communications engineering. His research
interests lie in communications and signal processing, areas in
which he has published about 180 papers in refereed journals and
conferences. Dr. de Lamare serves as associate editor for the
EURASIP Journal on Wireless Communications and Networking. He is a
Senior Member of the IEEE has served as the General Chair of the
7th IEEE International Symposium on Wireless Communications
Systems, held in York, UK in September 2010.
\end{biography}

\begin{biography}{Raimundo Sampaio-Neto}
received the Diploma and the M.Sc. degrees, both in electrical
engineering, from Pontificia Universidade Católica do Rio de
Janeiro (PUC-Rio) in 1975 and 1978, respectively, and the Ph.D.
degree in electrical engineering from the University of Southern
California (USC), Los Angeles, in 1983. From 1978 to 1979 he was
an Assistant Professor at PUC-Rio, and from 1979 to 1983 he was a
doctoral student and a Research Assistant in the Department of
Electrical Engineering at USC with a fellowship from CAPES. From
November 1983 to June 1984 he was a Post-Doctoral fellow at the
Communication Sciences Institute of the Department of Electrical
Engineering at USC, and a member of the technical staff of
Axiomatic Corporation, Los Angeles. He is now a researcher at the
Center for Studies in Telecommunications (CETUC) and an Associate
Professor of the Department of Electrical Engineering of PUC-Rio,
where he has been since July 1984. During 1991 he was a Visiting
Professor in the Department of Electrical Engineering at USC.
Prof. Sampaio has participated in various projects and has
consulted for several private companies and government agencies.
He was co-organizer of the Session on Recent Results for the IEEE
Workshop on Information Theory, 1992, Salvador. He has also served
as Technical Program co-Chairman for IEEE Global
Telecommunications Conference (Globecom'99) held in Rio de Janeiro
in December 1999 and as a member of the technical program
committees of several national and international conferences. He
was in office for two consecutive terms for the Board of Directors
of the Brazilian Communications Society where he is now a member
of its Advisory Council and Associate Editor of the Journal of the
Brazilian Communication Society. His areas of interest include
communication systems theory, digital transmission, satellite
communications and signal processing for communications.
\end{biography}

\begin{biography}{Martin Haardt}
(S'90 - M'98 - SM'99) has been a Full Professor in the Department
of Electrical Engineering and Information Technology and Head of
the Communications Research Laboratory at Ilmenau University of
Technology, Germany, since 2001. After studying electrical
engineering at the Ruhr-University Bochum, Germany, and at Purdue
University, USA, he received his Diplom-Ingenieur (M.S.) degree
from the Ruhr-University Bochum in 1991 and his Doktor-Ingenieur
(Ph.D.) degree from Munich University of Technology in 1996. In
1997 he joined Siemens Mobile Networks in Munich, Germany, where
he was responsible for strategic research for third generation
mobile radio systems. From 1998 to 2001 he was the Director for
International Projects and University Cooperations in the mobile
infrastructure business of Siemens in Munich, where his work
focused on mobile communications beyond the third generation.
During his time at Siemens, he also taught in the international
Master of Science in Communications Engineering program at Munich
University of Technology. Martin Haardt has received the 2009 Best
Paper Award from the IEEE Signal Processing Society, the Vodafone
(formerly Mannesmann Mobilfunk) Innovations-Award for outstanding
research in mobile communications, the ITG best paper award from
the Association of Electrical Engineering, Electronics, and
Information Technology (VDE), and the Rohde \& Schwarz Outstanding
Dissertation Award. In the fall of 2006 and the fall of 2007 he
was a visiting professor at the University of Nice in
Sophia-Antipolis, France, and at the University of York, UK,
respectively. His research interests include wireless
communications, array signal processing, high-resolution parameter
estimation, as well as numerical linear and multi-linear algebra.
Prof. Haardt has served as an Associate Editor for the IEEE
Transactions on Signal Processing (2002-2006), the IEEE Signal
Processing Letters (2006-2010), the Research Letters in Signal
Processing (2007-2009), the Hindawi Journal of Electrical and
Computer Engineering (since 2009), and as a guest editor for the
EURASIP Journal on Wireless Communications and Networking. He has
also served as the technical co-chair of the IEEE International
Symposiums on Personal Indoor and Mobile Radio Communications
(PIMRC) 2005 in Berlin, Germany, and as the technical program
chair of the IEEE International Symposium on Wireless
Communication Systems (ISWCS) 2010 in York, UK.
\end{biography}

\end{document}